\def\Jl{\ton{\bds{\hat{J}}\bds\cdot\bds{\hat{l}}}}
\def\Jm{\ton{\bds{\hat{J}}\bds\cdot\bds{\hat{m}}}}

\def\nk{n_{\rm b}}

\def\Pb{P_{\rm b}}

\def\rfr#1{Equation\,(\ref{#1})}
\def\rfrs#1#2{Equations\,(\ref{#1})-(\ref{#2})}

\def\Rfrs#1#2{Equations\,(\ref{#1})-(\ref{#2})}
\def\derp#1#2{\rp{\partial{#1}}{\partial{#2}}}
\def\dert#1#2{\frac{{{\textrm{d}}}{#1}}{{{\textrm{d}}}{#2}}}
\def\dertt#1#2{\frac{{{\textrm{d}^2}}{#1}}{{{\textrm{d}}}{#2}}}

\def\virg#1{``#1"}

\def\eqi{\begin{equation}}
\def\eqf{\end{equation}}
\def\eqia{\begin{eqnarray}}
\def\eqfa{\end{eqnarray}}

\def\rp#1#2{{#1\over#2}}
\def\lb#1{\label{#1}}

\def\bds#1{\boldsymbol{#1}}

\def\ton#1{\left(#1\right)}
\def\qua#1{\left[#1\right]}
\def\grf#1{\left\{#1\right\}}
\def\ang#1{\left\langle #1\right\rangle}
\documentclass[onecolumn]{aastex}
\usepackage{morefloats}
\usepackage[title]{appendix}
\usepackage{booktabs}
\usepackage[table,xcdraw]{xcolor}
\usepackage{multirow}
\usepackage{rotating,tabularx}
\usepackage{float}
\usepackage{enumerate}
\usepackage{rotating}
\usepackage[polutonikogreek,english]{babel}
\usepackage{amsmath,starfont,textgreek,w-greek,wasysym}
\usepackage{amsthm}
\usepackage{amscd,lineno}
\usepackage{amssymb,dsfont}
\usepackage{graphicx,epsfig}
\usepackage{txfonts}
\usepackage{xr-hyper}
\usepackage{hyperref}
\RequirePackage{color}

\newcommand{\emaila}{lorenzo.iorio@libero.it}

\allowdisplaybreaks[1]

\begin{document}

\title{Is there still something left that Gravity Probe B can measure?}

\shortauthors{L. Iorio}

\author{Lorenzo Iorio\altaffilmark{1} }
\affil{Ministero dell'Istruzione, dell'Universit\`{a} e della Ricerca
(M.I.U.R.)-Istruzione
\\ Permanent address for correspondence: Viale Unit\`{a} di Italia 68, 70125, Bari (BA),
Italy}

\email{\emaila}

\begin{abstract}
We perform a full analytical and numerical treatment, to the first post-Newtonian (1pN) order, of the general relativistic long-term spin precession of an orbiting gyroscope due to the mass quadrupole moment $J_2$ of its primary without any restriction on either the gyro's orbital configuration and the orientation in space of the symmetry axis $\bds{\hat{k}}$ of the central body. We apply our results to the past spaceborne Gravity Probe B (GP-B) mission by finding a secular rate of its spin's declination $\delta$ which may be as large as $\lesssim 30-40\,\mathrm{milliarcseconds\,per\,year\,\ton{\mathrm{mas\,yr}^{-1}}}$, depending on the initial orbital phase $f_0$. Both our analytical calculation and our simultaneous integration of the equations for the parallel transport of the spin 4-vector $\textsf{\textbf{\textsl{S}}}$ and of the geodesic equations of motion of the gyroscope confirm such a finding. For GP-B, the reported  mean error in measuring the spin's declination rate amounts to $\sigma^\mathrm{GP-B}_{\dot\delta}=18.3\,\mathrm{mas\,yr}^{-1}$. We also calculate  the general analytical expressions of the gravitomagnetic spin precession induced by the primary's angular momentum $\bds J$. In view of their generality, our results can be extended also to other astronomical and astrophysical scenarios of interest like, e.g., stars orbiting galactic supermassive black holes, exoplanets close to their parent stars, tight binaries hosting compact stellar corpses.
\end{abstract}

keywords{
general relativity and gravitation; experimental studies of gravity; experimental tests of
gravitational theories; satellite orbits; harmonics of the gravity potential field
}
\section{Introduction}
To the first post-Newtonian (1pN) level of order $\mathcal{O}\ton{c^{-2}}$, where $c$ is the speed of light in vacuum, the geodesic motion of a test particle through the deformed spacetime outside an axially symmetric oblate body of mass $M$, equatorial radius $R$ and dimensionless mass quadrupole moment $J_2$ is characterized by certain secular orbital precessions \citep{1988CeMec..42...81S,Sof89,1991ercm.book.....B,2015IJMPD..2450067I}. They have recently gained attention, being possibly detectable in some proposed space-based experiments like, e.g., HERO \citep{2019Univ....5..165I}.

Here, we will look at the long-term 1pN rate of change, proportional to $J_2\,c^{-2}$, of the spin $\bds{S}$ of a pointlike gyroscope freely moving with velocity $\bds v$ around an oblate primary. The analogous 1pN gyro's precessional effects due to only the mass monopole (the mass $M$) and the spin dipole (the proper angular momentum $\bds J$) moments of the central body acting as source of the gravitational field are the time-honored de Sitter-Fokker (or geodetic) \citep{1916MNRAS..77..155D,1921KNAB...23..729F} and Pugh-Schiff \citep{Pugh59,Schiff60} precessions, respectively. They were recently measured by the spaceborne  mission Gravity Probe B (GP-B)  in the field of Earth to $\simeq 0.3\%$ and $\simeq 19\%$, respectively, \citep{2011PhRvL.106v1101E,2015CQGra..32v4001E}, despite a higher accuracy had been originally expected \citep{Varenna74,2001LNP...562...52E}. We will not restrict ourselves to any particular orbital configuration of the moving gyroscope, and the symmetry axis of the oblate primary will retain an arbitrary orientation in space. We will calculate the sought effect both numerically and analytically by finding, among other things, that it depends on the initial position of the gyro along its orbit. In the case of GP-B, it turns out that the rate of change of the spin's declination (DEC) $\delta$, averaged over an orbital revolution,  may be as large as $\simeq 30-40\,\mathrm{milliarcseconds\,per\,year\,\ton{\mathrm{mas\,yr}^{-1}}}$. Thus, it may be potentially measurable in a future data reanalysis since the reported average experimental accuracy in measuring the temporal evolution of 
$\delta$ is \citep{2011PhRvL.106v1101E,2015CQGra..32v4001E} $\sigma^\mathrm{GP-B}_\mathrm{\dot\delta}=18.3\,\mathrm{mas\,yr}^{-1}$.
For previous analytical calculations, relying upon various simplifying assumptions concerning the gyro's orbit and different computational approaches, see
\citet{1969Ap&SS...4..119O,1970PhRvD...2.1428B,1988nznf.conf..685B,2003nlgd.conf..145A}. Even putting aside the issue of the particular orbital configurations adopted, they are, at least, incomplete since they neglect an important feature in the averaging procedure yielding to the dependence on the gyro's initial conditions which, instead, we will  take into account. Our simultaneous numerical integrations of the equations of motion of the gyro and of its spin will display it, by supporting our analytical findings. Moreover, it seems that the aforementioned works return incorrect results even for the part which is independent of the initial conditions, being also in mutual disagreement. In the following, we will not deal too much with the spin's right ascension (RA)  $\alpha$ since it turns out that, for GP-B, its total rate of change of the order of $\mathcal{O}\ton{J_2\,c^{-2}}$ is negligible.

For the sake of completeness, we will analytically derive also the generalization of the Pugh-Schiff gravitomagnetic spin precession valid for an arbitrary orientation of the primary's angular momentum $\bds J$ and for a generic orbital configuration of the gyroscope.

The generality of our approach allows our results to be extended also to other astronomical and astrophysical scenarios of interest like, e.g., other planets of our solar system, exoplanets, binaries with compact stellar corpses, supermassive black holes orbited by planets and stars. To this aim, it may be interesting to recall that \citet{1975Ap&SS..32....3H} investigated the possibility of using spacecraft-based missions to measure the  angular momenta of Jupiter an the Sun by means of the gravitomagnetic Pugh-Schiff spin precession.

The outline of the paper is as follows.
In Section\,\ref{numero}, we  numerically calculate the total spin precession of the order of $\mathcal{O}\ton{J_2\,c^{-2}}$ by simultaneously integrating the equations for the parallel transport of the gyro's spin 4-vector and the geodesic equations of  motion of the gyroscope. The  spin and orbital configurations of GP-B are used. Section\,\ref{pippo} is devoted to the analytical calculation. It, first, includes the direct effects (Section\,\ref{sub2.1}), which are the de Sitter precession for an arbitrary orbital configuration (Section\,\ref{sec3.1}), and the component of the spin rate of change  of the order of $\mathcal{O}\ton{J_2\,c^{-2}}$ arising from using a fixed Keplerian ellipse for the orbital average  (Section\,\ref{sec3.2}). Then, in Section\,\ref{sub2.2}, we calculate the indirect, or mixed, components of the sought precession. They are those arising from averaging the instantaneous 1pN de Sitter-like spin rate over the orbital period of a $J_2$-driven precessing ellipse (Section\,\ref{sec4.1}), and those coming from the inclusion   of the instantaneous orbital shifts caused by $J_2$ in the averaging procedure (Section\,\ref{sec4.2}). The total analytical spin precession of the order of $\mathcal{O}\ton{J_2\,c^{-2}}$ is discussed in Section\,\ref{totJ2}, where the GP-B case is illustrated and compared with the numerical results of Section\,\ref{numero}. The general expression of the gravitomagnetic spin precession is analytically calculated in Section\,\ref{gravimat}. Section\,\ref{fine} summarizes our finding and offers our conclusions.
\section{Numerical simulations: simultaneously integrating the equations for the motion of the gyroscope and of its spin}\lb{numero}
The equations for the parallel transport of the spin 4-vector $\textsf{\textbf{\textsl{S}}}$ of a pointlike gyroscope freely moving in the deformed spacetime of a central body are \citep{2013Zee,2013grsp.book.....O,2017grav.book.....M,2018tegp.book.....W}
\eqi
\dert{\textsf{\textsl{S}}^\nu}{\tau} = -\Gamma^\nu_{\lambda\beta}\,\textsf{\textsl{S}}^\lambda\,\dert{x^\beta}{\tau},\,\nu=0,\,1,\,2,\,3,\lb{trp}
\eqf
where $\tau$ is the gyro's proper time,
\eqi
\Gamma^\nu_{\lambda\beta}=\rp{1}{2}g^{\nu\gamma}\ton{\derp{g_{\gamma\lambda}}{x^\beta} +  \derp{g_{\gamma\beta}}{x^\lambda} - \derp{g_{\lambda\beta}}{x^\gamma}  },\,\nu,\,\lambda,\,\beta=0,\,1,\,2,\,3
\eqf
are the spacetime's Christoffel symbols, $g_{\gamma\lambda},\,g^{\gamma\lambda},\,\gamma,\,\lambda=0,\,1,\,2,\,3$ are the components of the spacetime metric tensor and of its inverse, respectively, and $dx^\beta/d\tau,\,\beta=0,\,1,\,2,\,3$ are the components of the gyro's 4-velocity $\textsf{\textbf{\textsl{u}}}$. The space-like components of $\textsf{\textbf{\textsl{S}}}$ are the components of the gyro's  spin vector $\bds S$, i.e. $\textsf{\textsl{S}}^i=S^i,\,i=1,\,2,\,3$. The time-like component $\textsf{\textsl{S}}^0$ of $\textsf{\textbf{\textsl{S}}}$ is determined by the constraint
\footnote{It is so because, in the gyro's rest frame, $\textsf{\textbf{\textsl{S}}}$ is space-like, while $\textsf{\textbf{\textsl{u}}}$ is time-like; thus, they are orthogonal.}
\eqi
g_{\rho\sigma}\,\textsf{\textsl{S}}^\rho\,\dert{x^\sigma}{\tau}=0.\lb{condiz}
\eqf

The geodesic equations of motion of the pointlike gyroscope are
\eqi
\dertt{x^\nu}{\tau^2} = -\Gamma^\nu_{\lambda\beta}\,\dert{x^\lambda}{\tau}\,\dert{x^\beta}{\tau},\,\nu=0,\,1,\,2,\,3.\lb{geod}
\eqf

In standard pN isotropic coordinates, the components  of the metric tensor of the spacetime of an isolated body are \citep{Sof89,2014grav.book.....P}
\begin{align}
g_{00} \lb{g00}& = 1 + \rp{2\,U}{c^2} + \mathcal{O}\ton{c^{-4}}, \\ \nonumber \\
g_{0i}&= \mathcal{O}\ton{c^{-3}},\,i=1,\,2,\,3,\\ \nonumber \\
g_{ij} \lb{gii}&= -\delta_{ij}\,\ton{1-\rp{2\,U}{c^2}}+ \mathcal{O}\ton{c^{-4}},\,i,\,j=1,\,2,\,3,
\end{align}
where
\eqi
\delta_{ij}=\left\{
              \begin{array}{ll}
                1 & \hbox{for $i=j$} \\
                0 & \hbox{for $i\neq j$,}
              \end{array}
            \right.
\,i,\,j=1,\,2,\,3
\eqf
is the Kronecker delta.
In the following, we will use cartesian coordinates, so that $x^1=x,\,x^2=y,\,x^3=z,\,S^1=S_x,\,S^2=S_y,\,S^3=S_z$.
In \rfrs{g00}{gii}, the potential $U\ton{\bds r}$ of the oblate mass is
\eqi
U = -\rp{\mu}{r}\qua{1-J_2\,\ton{\rp{R}{r}}^2\,\mathcal{P}_2\ton{\xi}}.\lb{Upot}
\eqf
In \rfr{Upot}, $\mu\doteq GM$ is the gravitational parameter of the central body, $G$ is the Newtonian constant of gravitation,
\eqi
\mathcal{P}_2\ton{\xi}=\rp{3\,\xi^2-1}{2}
\eqf
is the Legendre polynomial of degree 2, while
\eqi
\xi\doteq\bds{\hat{k}}\bds\cdot\bds{\hat{r}}
\eqf
is the cosine of the angle  between the body's symmetry axis $\bds{\hat{k}}$ and the unit position vector $\bds{\hat{r}}$.
In the case of a diagonal metric, as for \rfrs{g00}{gii}, \rfr{condiz} yields
\eqi
\textsf{\textsl{S}}^0 = -\rp{1}{c\,g_{00}}\ton{S_x\,g_{11}\,\dert{x}{\tau}+S_y\,g_{22}\,\dert{y}{\tau}+S_z\,g_{33}\,\dert{z}{\tau}}.\lb{S0}
\eqf

We set up a numerical code to simultaneously integrate both \rfr{trp} and \rfr{geod} for an arbitrary orientation of $\bds{\hat{k}}$ in space and unrestricted orbital configurations for the moving gyroscope. The space-like components of $\textsf{\textbf{\textsl{S}}}$ are parameterized in terms of two spherical angles $\alpha,\,\delta$ as
\begin{align}
S_x \lb{sx} & = \cos\delta\,\cos\alpha, \\ \nonumber \\
S_y & =\cos\delta\,\sin\alpha, \\ \nonumber \\
S_z \lb{sz} & =\sin\delta,
\end{align}
which, in the case of Earth and an equatorial coordinate system, are the spin's right ascension and declination, respectively.
As initial conditions for both the gyroscope orbit and its spin, we adopt those of GP-B \citep{TabGPB}, summarized in Table\,\ref{tavola1}.
As far as the initial value of $\textsf{\textsl{S}}^0$  is concerned, it can be retrieved from the condition of \rfr{condiz}. The  initial values of the space-like components of the 4-velocity $\textsf{\textbf{\textsl{u}}}$ can be obtained from
\eqi
\textsf{\textsl{u}}^i = \dert{t}{\tau}\,v^i,\,i=1,\,2,\,3,
\eqf
where
$v^i,\,i=1,\,2,\,3$ are the components of the velocity $\bds v$ (see \rfr{vu}), and
\eqi
\dert{t}{\tau} = c\,\sqrt{\rp{1}{g_{\rho\sigma}\,\dert{x^\rho}{t}\,\dert{x^\sigma}{t}}}.
\eqf
\begin{table}[ht]
\caption{Initial conditions common to all the numerical integrations. They were retrieved from \citet{TabGPB} for GP-B. The true anomaly at epoch $f_0$ is changed from one run to another.}
\label{tavola1}
\begin{center}
\begin{tabular}{|l|l|l|l|}
\hline
Orbital and spin parameter & Symbol & Value & Unit\\
\hline
Semimajor axis & $a$ & $7027.4$ & km \\
Eccentricity & $e$ & $0.0014$ & - \\
Inclination & $I$ & $90.007$ & deg \\
Longitude of the ascending node & $\mathit{\Omega}$ & $163.26$ & deg\\
Argument of perigee & $\omega$ & $71.3$ & deg \\
True anomaly at epoch & $f_0$ & variable & deg \\
DEC of the spin axis & $\delta$ & $0$ & deg \\
RA of the spin axis & $\alpha$ & $\mathit{\Omega}+180^\circ$ & deg\\
\hline
\end{tabular}
\end{center}
\end{table}

\begin{figure}[htb!]
\begin{center}
\centerline{
\vbox{
\begin{tabular}{c}
\epsfysize= 8.0 cm\epsfbox{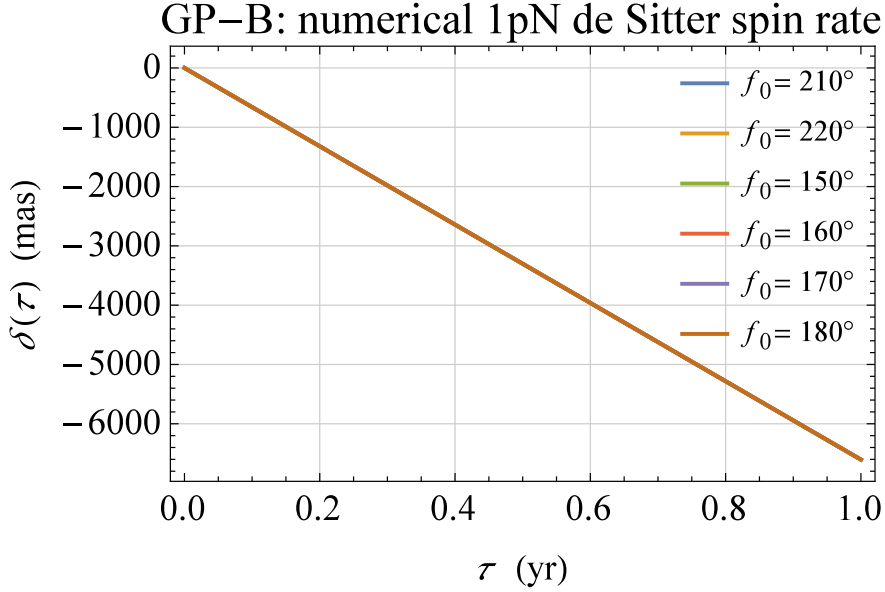}\\
\end{tabular}
}
}
\caption{
Numerically produced 1pN de Sitter time series $\delta\ton{\tau}$, in mas, of the declination $\delta$,  of the spin axis of a gyroscope orbiting the Earth along a Keplerian ellipse for some different initial values of the true anomaly $f_0$. Each of the times series was obtained by simultaneously integrating \rfr{trp} and \rfr{geod} with $J_2=0$ in \rfrs{g00}{Upot}, and calculating $\arcsin S_z\ton{\tau}$ for  the resulting solution $S_z\ton{\tau}$ of each run. It turned out that essentially the same outcome can also be obtained by replacing \rfr{geod} with the 3-dimensional Newtonian acceleration ${\bds{A}}_\mathrm{N} = -\ton{\mu/r^2}\,\bds{\hat{r}}$ for the gyroscope. The initial conditions adopted, common to all of the integrations, were those of GP-B \citep{TabGPB}, summarized in Table\ref{tavola1}. All the shifts are independent of $f_0$, and their slopes amount just to $\dot\delta_\mathrm{dS} = -6603.8\,\mathrm{mas\,yr}^{-1}$, as expected.}\label{figura0}
\end{center}
\end{figure}

We, first, test our routine by successfully reproducing the de Sitter precession, shown in Figure\,\ref{figura0}.
The time series in it were obtained by switching off $J_2$ in both \rfr{trp} and \rfr{geod}. They correspond to the orbital average  over a Keplerian ellipse\footnote{\lb{foot1}In fact, even switching off $J_2$ in \rfr{geod} does not correspond to a purely Keplerian path but to a (slowly) precessing ellipse affected, to the 1pN level, by a perigee precession analogous to the Einstein precession of the perihelion of Mercury. However, its impact on the spin precession is negligible, being of the order of $\mathcal{O}\ton{c^{-4}}$.} of the 1pN components of the right-hand-sides of \rfr{trp} for $i=1,\,2,\,3$ and $J_2=0$.   As expected, all the signatures in Figure\,\ref{figura0} are independent of $f_0$.

\begin{figure}[htb!]
\begin{center}
\centerline{
\vbox{
\begin{tabular}{c}
\epsfysize= 8.0 cm\epsfbox{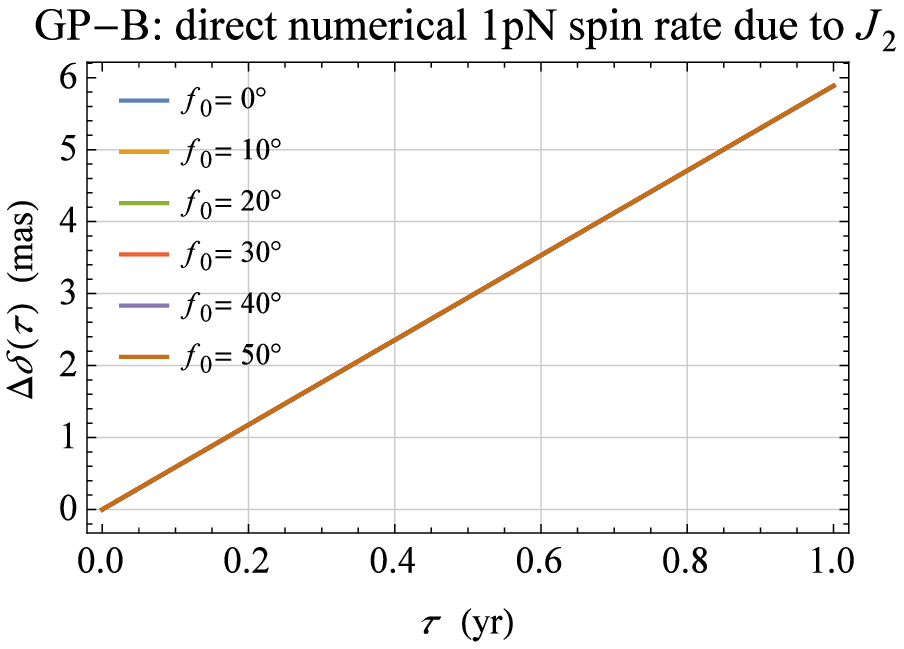}\\
\end{tabular}
}
}
\caption{
Numerically produced direct 1pN $J_2$-induced yearly shifts $\Delta\delta\ton{\tau}$, in mas, of the declination $\delta$ of the spin axis of a gyroscope orbiting the Earth along a fictitious  Keplerian ellipse for some different initial values of the true anomaly $f_0$. Each of the times series $\Delta\delta\ton{\tau}$ was obtained by simultaneously integrating \rfr{trp} with $J_2\neq 0$ and \rfr{geod} with $J_2=0$, and subtracting from each of them the corresponding time series obtained by integrating both \rfr{trp} and \rfr{geod} with $J_2=0$ (the de Sitter trends). Then, $\arcsin S_z\ton{\tau}$ was calculated for each of the resulting solutions $S_z\ton{\tau}$. The initial conditions adopted, common to all of the runs, were those of GP-B \citep{TabGPB}, summarized in Table\,\ref{tavola1}. All the shifts are independent of $f_0$, and agree with the result calculated analytically in Section\,\ref{sec3.2}. The slope amounts to $\dot\Delta\delta=5.8\,\mathrm{mas\,yr}^{-1}$. }\label{figura1}
\end{center}
\end{figure}

Figure\,\ref{figura1} displays the \virg{direct} part of the spin precession of the order of $\mathcal{O}\ton{J_2\,c^{-2}}$ obtained by restoring $J_2$ in \rfr{trp}, but not in \rfr{geod}, and subtracting from the resulting signatures the purely de Sitter ones.
It essentially corresponds to the orbital average of the 1pN components of the right-hand-sides of \rfr{trp} for $i=1,\,2,\,3$ and $J_2\neq 0$  over an actually non-existent Keplerian ellipse\footnote{If $J_2\neq 0$, the actual trajectory is a (slowly) precessing ellipse \citep{2005som..book.....C}. See also Footnote\,\ref{foot1}.}. Clearly, it is an unphysical situation which is just an intermediate check of our analytical calculation, to be displayed in Section\,\ref{sec3.2}, and of the results existing in the literature.  Its slope amounts to $5.8\,\mathrm{mas\,yr}^{-1}$, and is independent of $f_0$. As we will see in Section\,\ref{sec3.2}, our analytical outcome for the direct precession of the order of $\mathcal{O}\ton{J_2\,c^{-2}}$ agrees with Figure\,\ref{figura1} to within $\simeq 0.6\,\mathrm{mas\,yr}^{-1}$.
Instead, the part of Equation\,(53) of \citet{1970PhRvD...2.1428B} containing $J_2$ allows to obtain $\dot\Delta\delta = 4.1\,\mathrm{mas\,yr}^{-1}$, while the $J_2$-dependent part of $\ang{\mathrm{\Omega}_G}$ in \citet[pag.\,153]{2003nlgd.conf..145A} corresponds to  $\dot\Delta\delta = 6.6\,\mathrm{mas\,yr}^{-1}$. As it will be demonstrated in Section\,\ref{sec3.2}, both of them disagree with our analytical calculation.

\begin{figure}[htb!]
\begin{center}
\centerline{
\vbox{
\begin{tabular}{cc}
\epsfysize= 5.4 cm\epsfbox{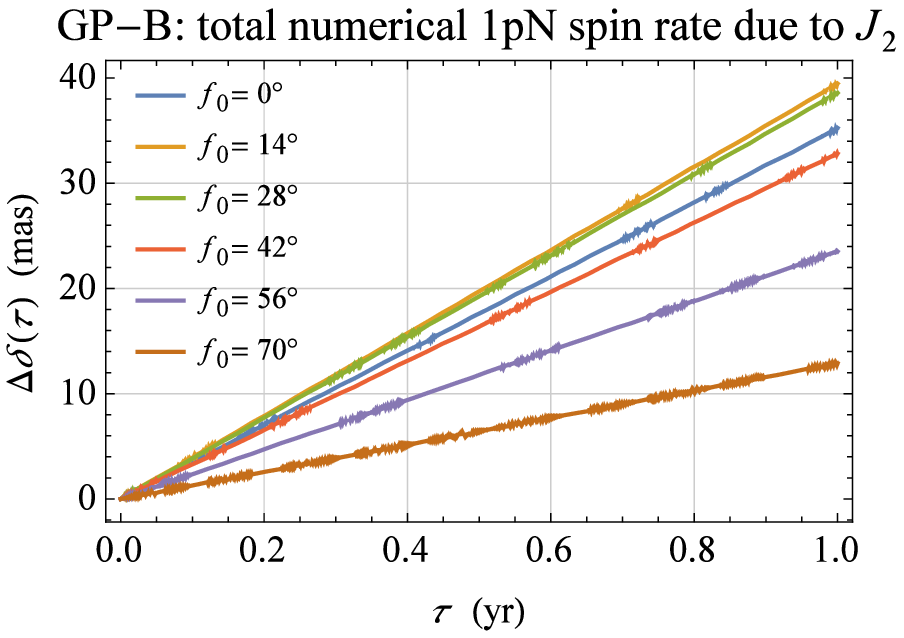} & \epsfysize= 5.4 cm\epsfbox{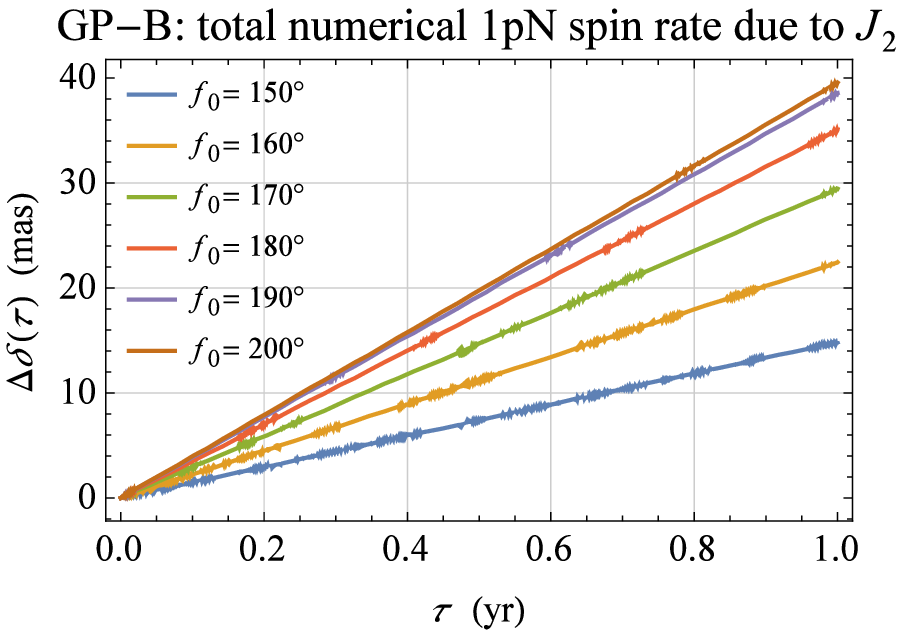}\\
\epsfysize= 5.4 cm\epsfbox{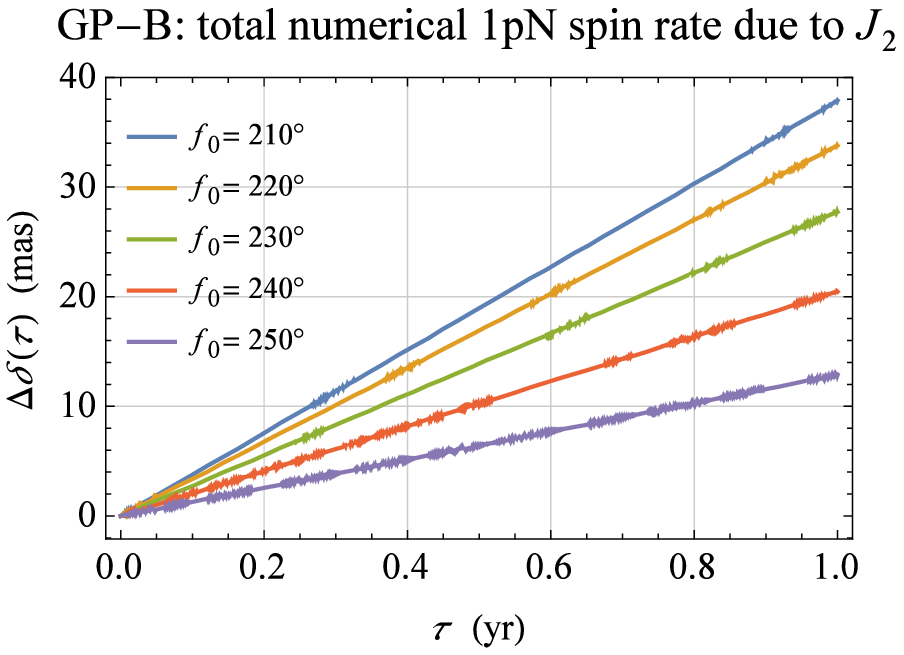} & \epsfysize= 5.4 cm\epsfbox{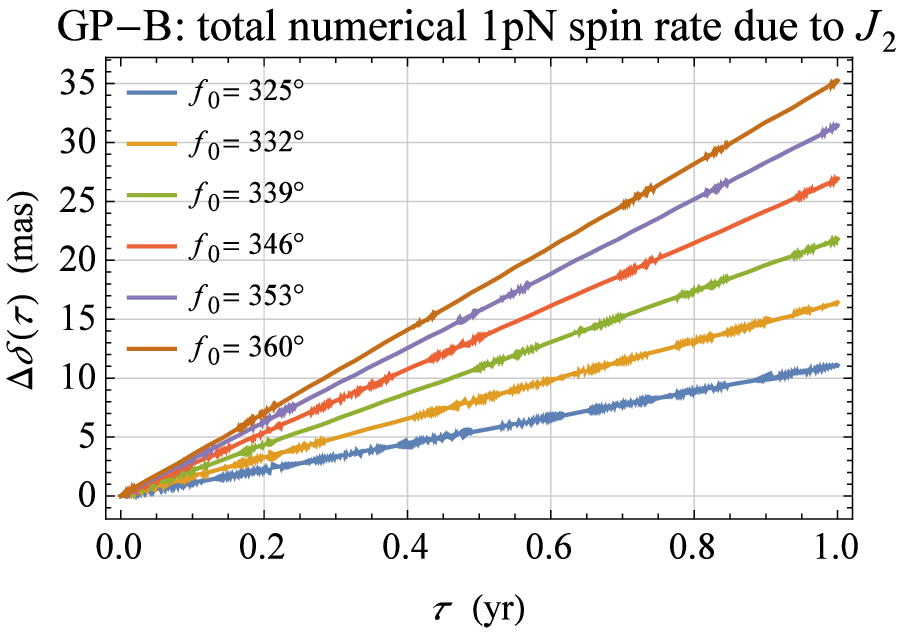}\\
\end{tabular}
}
}
\caption{
Numerically produced full 1pN $J_2$-induced yearly shifts $\Delta\delta\ton{\tau}$ of the declination $\delta$  of the spin axis of a gyroscope orbiting the oblate Earth along a realistic precessing Keplerian ellipse for different values of the true anomaly $f_0$. Each of the time series $\Delta\delta\ton{\tau}$ was obtained by simultaneously integrating \rfr{trp} and \rfr{geod}, both with $J_2\neq 0$ in \rfrs{g00}{Upot} and subtracting the corresponding de Sitter trends from each of them, and calculating $\arcsin S_z\ton{\tau}$ for
the resulting solution $S_z\ton{\tau}$ of each run. The initial conditions adopted, common to all of the integrations, were those of GP-B \citep{TabGPB}, summarized in Table\,\ref{tavola1}. All the numerically integrated shifts agree with those calculated analytically in Section\,\ref{totJ2} to within $\lesssim 5-8\, \mathrm{mas\,yr}^{-1}$. Such a discrepancy is not statistically significative since it is smaller than $\sigma^\mathrm{GP-B}_{\dot\delta}=18.3\,\mathrm{mas\,yr}^{-1}$ \citep{2011PhRvL.106v1101E,2015CQGra..32v4001E}.
}\label{figura2}
\end{center}
\end{figure}

The total spin precessions of the order of $\mathcal{O}\ton{J_2\,c^{-2}}$, obtained by simultaneously integrating both \rfr{trp} and \rfr{geod} with $J_2\neq 0$  for different values of $f_0$ and subtracting the purely de Sitter trends from the resulting signatures, are displayed in Figure\,\ref{figura2}.
They can be thought as the sum of the direct precession of Figure\,\ref{figura1} and of the \virg{indirect}, or \virg{mixed}, ones arising from the fact that, in this case, the trajectory of the gyroscope is, more realistically, a (slowly) precessing ellipse mainly driven by\footnote{See Footnote\,\ref{foot1}.} $J_2$. It can be thought as if, in addition to the Keplerian average of the $J_2$-dependent parts of the space-like components of \rfr{trp}, the de Sitter-like 1pN components of the right-hand-sides of \rfr{trp} for $i=1,\,2,\,3$ and $J_2=0$ were averaged over one orbital revolution by taking now into account also the $J_2$-induced instantaneous changes of the osculating Keplerian orbital elements parameterizing the varying ellipse, and  the fact that the orbital period is the time interval between two successive passages at a changing perigee. The same, in principle, would hold also for the 1pN orbital changes which, however, would affect the spin precession to the $1/c^4$ level. Effects of the order of $\mathcal{O}\ton{J_2^2\,c^{-2}}$ would arise by repeating the same average for the  1pN components of the right-hand-sides of \rfr{trp} for $i=1,\,2,\,3$ and $J_2\neq 0$. Our numerical integration accounts simultaneously for all such negligible effects of higher order as well. A striking feature of Figure\,\ref{figura2} is that the indirect effects induce a neat dependence on $f_0$ which can yield spin precessions as large as $\simeq 30-40\,\mathrm{mas\,yr}^{-1}$. It is a quite important finding since the reported mean error in measuring the spin's declination precession of GP-B is $\sigma^\mathrm{GP-B}_{\dot\delta} =18.3\,\mathrm{mas\,yr}^{-1}$ \citep{2011PhRvL.106v1101E,2015CQGra..32v4001E}, and it may prompt some reanalysis of the mission data. Such a dependence on $f_0$ induced by the mixed effects is captured and reproduced by our analytical calculation of the overall precession in Sections\,\ref{sec3.2} to \ref{sec4.2} to within $\lesssim 5-8\,\mathrm{mas\,yr}^{-1}$; cfr. with Figure\,\ref{figura3} in Section\,\ref{totJ2}. Instead, it is missing in the literature. Indeed, if, on the one hand, \citet{2003nlgd.conf..145A} seemingly dealt only with the direct $J_2$-induced precession, on the other hand, \citet{1970PhRvD...2.1428B} were aware of such an issue, but they somehow treated it only partly since their Equation\,(52) does not contain any dependence on the initial orbital phase. Should it ever be related to the aforementioned issue of the orbital period in a precessing orbit, it is in disagreement with our analytical results for it, as we will show in Section\,\ref{sec4.1}.
\section{Analytical calculation}\lb{pippo}
By expanding
\eqi
\dert{\textsf{\textsl{S}}^i}{\tau} = -\Gamma^i_{\alpha\beta}\,\textsf{\textsl{S}}^\alpha\,\dert{x^\beta}{\tau},\,i=1,\,2,\,3,\lb{spr}
\eqf
calculated with \rfr{Upot} in \rfrs{g00}{gii},
to the order of $\mathcal{O}\ton{c^{-2}}$, one obtains the instantaneous rates of change of the gyro's spin components as
\begin{align}
\dert{S_x}{t} \lb{kaz} & = \ton{\textsf{T}^\mathrm{dS}_{xx}+\textsf{T}^{J_2}_{xx}}\,S_x + \ton{\textsf{T}^\mathrm{dS}_{xy}+\textsf{T}^{J_2}_{xy}}\,S_y + \ton{\textsf{T}^\mathrm{dS}_{xz}+\textsf{T}^{J_2}_{xz}}\,S_z, \\ \nonumber \\
\dert{S_y}{t} & = \ton{\textsf{T}^\mathrm{dS}_{yx}+\textsf{T}^{J_2}_{yx}}\,S_x + \ton{\textsf{T}^\mathrm{dS}_{yy}+\textsf{T}^{J_2}_{yy}}\,S_y + \ton{\textsf{T}^\mathrm{dS}_{yz}+\textsf{T}^{J_2}_{yz}}\,S_z, \\ \nonumber \\
\dert{S_z}{t} \lb{kuz} & = \ton{\textsf{T}^\mathrm{dS}_{zx}+\textsf{T}^{J_2}_{zx}}\,S_x + \ton{\textsf{T}^\mathrm{dS}_{zy}+\textsf{T}^{J_2}_{zy}}\,S_y + \ton{\textsf{T}^\mathrm{dS}_{zz}+\textsf{T}^{J_2}_{zz}}\,S_z,
\end{align}
where the coefficients of the matrices ${\textbf{\textsf{T}}}^\mathrm{dS},\,{\textbf{\textsf{T}}}^{J_2}$ are, in general, time-dependent. They are
\begin{align}
\textsf{T}^\mathrm{dS}_{xx} \lb{TxxdS} & = \rp{\mu}{c^2\,r^3}\,\ton{v_y\,y + v_z\,z}, \\ \nonumber \\
\textsf{T}^\mathrm{dS}_{xy} & = \rp{\mu}{c^2\,r^3}\,\ton{-2\,v_y\,x + v_x\,y}, \\ \nonumber \\
\textsf{T}^\mathrm{dS}_{xz} & = \rp{\mu}{c^2\,r^3}\,\ton{-2\,v_z\,x + v_x\,z}, \\ \nonumber \\
\textsf{T}^\mathrm{dS}_{yx} & = \rp{\mu}{c^2\,r^3}\,\ton{v_y\,x - 2\,v_x\,y}, \\ \nonumber \\
\textsf{T}^\mathrm{dS}_{yy} & = \rp{\mu}{c^2\,r^3}\,\ton{v_x\,x + v_z\,z}, \\ \nonumber \\
\textsf{T}^\mathrm{dS}_{yz} & = \rp{\mu}{c^2\,r^3}\,\ton{-2\,v_z\,y + v_y\,z}, \\ \nonumber \\
\textsf{T}^\mathrm{dS}_{zx} & = \rp{\mu}{c^2\,r^3}\,\ton{v_z\,x - 2\,v_x\,z}, \\ \nonumber \\
\textsf{T}^\mathrm{dS}_{zy} & =  \rp{\mu}{c^2\,r^3}\,\ton{v_z\,y - 2\,v_y\,z}, \\ \nonumber \\
\textsf{T}^\mathrm{dS}_{zz} \lb{TzzdS} & =  \rp{\mu}{c^2\,r^3}\,\ton{v_x\,x + v_y\,y},
\end{align}
and
\begin{align}
\textsf{T}^{J_2}_{xx} \lb{TxxJ2} & = \rp{3\,\mu\,J_2\,R^2}{2\,c^2\,r^7}\,\qua{3\,v_z\,\ton{x^2 + y^2}\,z - 2\,v_z\,z^3 + v_y\,y\,\ton{x^2 + y^2 - 4\,z^2}}, \\ \nonumber \\
\textsf{T}^{J_2}_{xy} & =  -\rp{3\,\mu\,J_2\,R^2}{2\,c^2\,r^7}\,\ton{2\,v_y\,x -\,v_x\,y}\,\ton{x^2 + y^2 - 4\,z^2}, \\ \nonumber \\
\textsf{T}^{J_2}_{xz} & =  \rp{3\,\mu\,J_2\,R^2}{2\,c^2\,r^7}\,\qua{3\,v_x\,\ton{x^2 + y^2}\,z - 2\,v_x\,z^3 - 2\,v_z\,x\,\ton{x^2 + y^2 - 4\,z^2}}, \\ \nonumber \\
\textsf{T}^{J_2}_{yx} & = \rp{3\,\mu\,J_2\,R^2}{2\,c^2\,r^7}\,\ton{v_y\,x - 2\,v_x\,y}\,\ton{x^2 + y^2 - 4\,z^2}, \\ \nonumber \\
\textsf{T}^{J_2}_{yy} & =  \rp{3\,\mu\,J_2\,R^2}{2\,c^2\,r^7}\,\qua{3\,v_z\,\ton{x^2 + y^2}\,z - 2\,v_z\,z^3 + v_x\,x\,\ton{x^2 + y^2 - 4\,z^2}}, \\ \nonumber \\
\textsf{T}^{J_2}_{yz} & =  \rp{3\,\mu\,J_2\,R^2}{2\,c^2\,r^7}\,\qua{3\,v_y\,\ton{x^2 + y^2}\,z - 2\,v_y\,z^3 - 2\,v_z\,y\,\ton{x^2 + y^2 - 4\,z^2}}, \\ \nonumber \\
\textsf{T}^{J_2}_{zx} & = \rp{3\,\mu\,J_2\,R^2}{2\,c^2\,r^7}\,\qua{-6\,v_x\,\ton{x^2 + y^2}\,z + 4\,v_x\,z^3 + v_z\,x\,\ton{x^2 + y^2 - 4\,z^2}}, \\ \nonumber \\
\textsf{T}^{J_2}_{zy} & =  \rp{3\,\mu\,J_2\,R^2}{2\,c^2\,r^7}\,\qua{-6\,v_y\,\ton{x^2 + y^2}\,z + 4\,v_y\,z^3 + v_z\,y\,\ton{x^2 + y^2 - 4\,z^2}}, \\ \nonumber \\
\textsf{T}^{J_2}_{zz} \lb{TzzJ2} & =  \rp{3\,\mu\,J_2\,R^2}{2\,c^2\,r^7}\,\ton{v_x\,x + v_y\,y}\,\ton{x^2 + y^2 - 4\,z^2}.
\end{align}
As far as the rates of change of the spin's spherical angles $\alpha,\,\delta$ are concerned, from \rfrs{sx}{sz} one gets
\begin{align}
\dert{\delta}t & = \lb{DEC} \rp{1}{\cos\delta}\dert{S_z}t, \\ \nonumber \\
\ton{\dert{\alpha}t}^2 & = \lb{RA}\rp{1}{\cos^2\delta}\qua{\ton{\dert{S_x}t}^2 + \ton{\dert{S_y}t}^2 - \tan^2\delta\,\ton{\dert{S_z}t}^2}.
\end{align}

Since we are interested in the long-term rate of change of $\bds S$, we must properly average the right-hand-sides of \rfrs{kaz}{kuz} over one orbital period $\Pb$. It requires care, especially for the effects of the order of $\mathcal{O}\ton{J_2\,c^{-2}}$.  Indeed, the actual orbital path of the gyroscope around its distorted primary is a generally slowly precessing ellipse \citep{2005som..book.....C}, not a fixed Keplerian one as it would be if it were\footnote{See Footnote\,\ref{foot1}.} $J_2=0$. This implies that, during an orbital revolution, all the Keplerian orbital elements characterizing the shape, the size and the orientation of the ellipse undergo instantaneous variations due to $J_2$ which should be taken into account in the averaging procedure since they give rise to effects which are just of the order of\footnote{In principle, also the spin components entering linearly the right-hand-sides of \rfrs{kaz}{kuz} do vary instantaneously. Nonetheless, since their changes are of the order of $\mathcal{O}\ton{c^{-2}}$ due to the de Sitter precession, they can be neglected in the average since they would affect the spin rates to the order of $\mathcal{O}\ton{c^{-4}}$.  } $\mathcal{O}\ton{J_2\,c^{2}}$. Moreover, the fact that the line of the apsides, from which the time-dependent true anomaly\footnote{The true anomaly yields the instantaneous position of the test particle along its orbit.} $f$ is reckoned, does vary during the orbital motion because of $J_2$ has to be taken into account as well, yielding further contributions of the order of $\mathcal{O}\ton{J_2\,c^{2}}$. Such \virg{indirect}, or \virg{mixed}, features are to be added to the direct ones arising from a straightforward average of \rfrs{TxxJ2}{TzzJ2} over an unperturbed Keplerian ellipse assumed as reference trajectory.

From a computational point of view, we can split the calculation of the averaged 1pN gyro's spin precession in two parts.
\subsection{The direct effects}\lb{sub2.1}
The first one deals with what one may define as the \virg{direct} effects, denoted in the following with the superscript $\virg{\mathrm{dir}}$, arising from averaging \rfrs{TxxdS}{TzzJ2}, evaluated onto an unchanging\footnote{It is considered, in the first instance, as fixed over a timescale comparable with the orbital period.} Keplerian ellipse. The latter is characterized by \citep{1991ercm.book.....B}
\begin{align}
p \lb{pi}&= a\ton{1-e^2}, \\ \nonumber \\
r & = \rp{p}{1+e\,\cos f}, \\ \nonumber \\
\dert t f \lb{deti} & = \rp{r^2}{\sqrt{\mu\,p}}, \\ \nonumber \\
\bds r \lb{erre}& = r\,\ton{\bds{\hat{P}}\,\cos f + \bds{\hat{Q}}\,\sin f},\\ \nonumber \\
\bds v \lb{vu}& = \sqrt{\rp{\mu}{p}}\,\qua{-\bds{\hat{P}}\,\sin f + \bds{\hat{Q}}\,\ton{\cos f + e}}.
\end{align}
In \rfrs{erre}{vu}, it is
\begin{align}
\bds{\hat{P}} & = \bds{\hat{l}}\,\cos\omega + \bds{\hat{m}}\,\sin\omega, \\ \nonumber \\
\bds{\hat{Q}} & = -\bds{\hat{l}}\,\sin\omega + \bds{\hat{m}}\,\cos\omega,
\end{align}
with
\begin{align}
\bds{\hat{l}} & = \grf{\cos\mathit{\Omega},\,\sin\mathit{\Omega},\,0}, \\ \nonumber \\
\bds{\hat{m}} \lb{emme}& = \grf{-\cos I\,\sin\mathit{\Omega},\,\cos I\,\cos\mathit{\Omega},\,\sin I}.
\end{align}
In \rfrs{pi}{emme}, $p,\,a\,,e\,,I\,,\mathit{\Omega}\,,\omega$ are the semilatus rectum, the semimajor axis, the eccentricity, the inclination, the longitude of the ascending node, and  the argument of pericentre, respectively, of the Keplerian ellipse. The size and the shape of the latter are fixed by $a$ and $e$, respectively. The inclination and the position of the orbital plane with respect to the reference $\grf{x,\,y}$ plane are determined by $I$ and $\mathit{\Omega}$, respectively; the line of the nodes is the intersection of the orbital plane with the reference $\grf{x,\,y}$ plane. The orientation of the ellipse within the orbital plane itself is characterized by $\omega$.  The unit vector $\bds{\hat{l}}$ is directed along the line of the nodes toward the ascending node, while $\bds{\hat{m}}$ lies in the orbital plane perpendicularly to $\bds{\hat{l}}$. The unit vector $\bds{\hat{P}}$ is directed along the line of the apsides toward the pericentre in the orbital plane where $\bds{\hat{Q}}$ stays transversely to $\bds{\hat{P}}$ itself.
Finally, we mention also the unit vector
\eqi
\bds{\hat{h}} = \grf{\sin I\,\sin\mathit{\Omega},\,-\sin I\,\cos\mathit{\Omega},\,\cos I},
\eqf
directed along the orbital angular momentum perpendicularly to the orbital plane\footnote{It turns out that $\bds{\hat{l}},\,\bds{\hat{m}},\bds{\hat{h}}$ are a right-handed triad of unit vectors.}.

The resulting direct effects consist of the usual de Sitter precession, and of one part of the 1pN spin's rate of change due to $J_2$. In  Section\,\ref{sec3.1} and Section\,\ref{sec3.2}, we will display the explicit expressions of the averaged matrix elements of \rfrs{TxxdS}{TzzJ2}. For the sake of simplicity, we will omit the brackets $\ang{\ldots}$ denoting the average over one orbital period throughout the paper.
\subsubsection{The de Sitter precession}\lb{sec3.1}
Let us introduce the following dimensional amplitude having the dimension of  reciprocal  time
\eqi
\mathcal{A}_\mathrm{dS}\doteq \rp{3}{4}\,\nk\,\ton{\rp{\mathcal{R}_\mathrm{s}}{a}}\,\rp{1}{\ton{1-e^2}},
\eqf
where $\mathcal{R}_\mathrm{s}\doteq 2\,\mu/c^2$ is the primary's Schwarzschild radius.
The analytical expressions of the average of \rfrs{TxxdS}{TzzdS} yield the geodetic precession for an arbitrary orbital configuration of the moving gyroscope.
We have
\begin{align}
\textsf{T}^{\mathrm{dS}}_{xx} \lb{fac}& = 0, \\ \nonumber \\
\textsf{T}^{\mathrm{dS}}_{xy} & = -\mathcal{A}_\mathrm{dS}\,\cos I, \\ \nonumber \\
\textsf{T}^{\mathrm{dS}}_{xz} & = -\mathcal{A}_\mathrm{dS}\,\sin I\,\cos\mathit{\Omega}, \\ \nonumber \\
\textsf{T}^{\mathrm{dS}}_{yx} & = \mathcal{A}_\mathrm{dS}\,\cos I, \\ \nonumber \\
\textsf{T}^{\mathrm{dS}}_{yy} & = 0, \\ \nonumber \\
\textsf{T}^{\mathrm{dS}}_{yz} & = -\mathcal{A}_\mathrm{dS}\,\sin I\,\sin\mathit{\Omega}, \\ \nonumber \\
\textsf{T}^{\mathrm{dS}}_{zx} \lb{zxdS}& = \mathcal{A}_\mathrm{dS}\,\sin I\,\cos\mathit{\Omega}, \\ \nonumber \\
\textsf{T}^{\mathrm{dS}}_{zy} & = \mathcal{A}_\mathrm{dS}\,\sin I\,\sin\mathit{\Omega}, \\ \nonumber \\
\textsf{T}^{\mathrm{dS}}_{zz} \lb{fuc}& = 0.
\end{align}
From \rfrs{DEC}{RA} and \rfrs{fac}{fuc}, it is possible to obtain
\begin{align}
\dert{\delta}{t} \lb{DES} & = \mathcal{A}_\mathrm{dS}\,\sin I\,\cos\ton{\alpha-\mathit{\Omega}}, \\ \nonumber \\
\ton{\dert{\alpha}{t}}^2 & = \mathcal{A}^2_\mathrm{dS}\,\ton{\cos I + \sin I\,\tan\delta\,\sin\ton{\alpha-\mathit{\Omega}}}^2.
\end{align}
Figure\,\ref{figura0} agrees with \rfr{DES} calculated for GP-B. \Rfrs{fac}{fuc} show that the 1pN spin rate due to the mass monopole of the primary  can be written as
\eqi
\dert{\bds{\hat{S}}}t = {\mathbf{\Omega}}_\mathrm{dS}\bds\times\bds{\hat{S}}\lb{megaDS},
\eqf
with
\eqi
{\mathbf{\Omega}}_\mathrm{dS}\doteq \mathcal{A}_\mathrm{dS}\,\bds{\hat{h}}.
\eqf
The vectorial expression of \rfr{megaDS} agrees with, e.g., (10.146a) of \citet{2014grav.book.....P} in the limit $e\rightarrow 0$.
\subsubsection{The  $J_2\,c^{-2}$ spin rate of change: direct part}\lb{sec3.2}
Let us introduce the following dimensional amplitude having the dimension of reciprocal time:
\eqi
{\mathcal{A}}_{J_2}\doteq \rp{\nk}{2}\,\ton{\rp{\mathcal{R}_\mathrm{s}}{a}}\,\ton{\rp{R}{a}}^2\,\rp{J_2}{\ton{1-e^2}^3}=\rp{2}{3}\,\ton{\rp{R}{a}}^2\,\rp{J_2}{\ton{1-e^2}^2}\,\mathcal{A}_\mathrm{dS}.
\eqf
In the following, we will display the averaged expressions of \rfrs{TxxJ2}{TzzJ2}. For the sake of simplicity, we will limit here to the case in which the reference $z$ axis is aligned with the unit vector $\bds{\hat{k}}$ of the body's symmetry axis. We have
\begin{align}
\textsf{T}^{J_2\,\mathrm{dir}}_{xx} \nonumber \lb{xxdirJ2} &  = -\rp{15}{512}\,{\mathcal{A}}_{J_2}\,\grf{-e^2\,\qua{\ton{5 + 20\,\cos 2I +
         7\,\cos 4I}\,\cos 2\mathit{\Omega} + \right.\right. \\ \nonumber \\
\nonumber &\left.\left. +
      4\,\ton{5 + 7\,\cos 2I}\,\sin^2 I}\,\sin 2\omega -
   4\,\qua{e^2\,\ton{3\,\cos I + 5\,\cos 3I}\,\cos 2\omega + \right.\right. \\ \nonumber \\
&\left.\left. +
      4\,\ton{2 + 3\,e^2}\,\cos I\,\sin^2 I}\,\sin 2\mathit{\Omega}}, \\ \nonumber \\
\textsf{T}^{J_2\,\mathrm{dir}}_{xy} \nonumber &= -\rp{3}{512}\,{\mathcal{A}}_{J_2}\,\grf{4\,\ton{3\,\cos I + 5\,\cos 3I}\,\ton{12 + 18\,e^2 + 5\,e^2\,\cos 2\omega\,\cos 2\mathit{\Omega}} + \right. \\ \nonumber \\
\nonumber &\left. +
   80\,\cos I\,\qua{9\,e^2\,\cos 2\omega +\,\ton{2 + 3\,e^2}\,\cos 2\mathit{\Omega}}\,\sin^2 I -\right. \\ \nonumber \\
&\left. -
   5\,e^2\,\ton{5 + 20\,\cos 2I + 7\,\cos 4I}\,\sin 2\omega\,\sin
     2\mathit{\Omega}}, \\ \nonumber \\
\textsf{T}^{J_2\,\mathrm{dir}}_{xz} \nonumber &= -\rp{15}{256}\,{\mathcal{A}}_{J_2}\,\grf{4\,\qua{6 + 9\,e^2 + \cos 2I\,\ton{10 + 15\,e^2 -
         4\,e^2\,\cos 2\omega}}\,\cos \mathit{\Omega}\,\sin I + \right. \\ \nonumber \\
&\left. + e^2\,\ton{22\,\sin 2I - 7\,\sin 4I}\,\sin 2\omega\,\sin \mathit{\Omega}}, \\ \nonumber \\
\textsf{T}^{J_2\,\mathrm{dir}}_{yx} \nonumber &= -\rp{3}{512}\,{\mathcal{A}}_{J_2}\,\grf{20\,\cos 3I\,\ton{-12 - 18\,e^2 +
      5\,e^2\,\cos 2\omega\,\cos 2\mathit{\Omega}} + \right. \\ \nonumber \\
\nonumber &\left. +
   4\,\cos I\,\qua{15\,e^2\,\cos 2\omega\,\ton{-6 + 6\,\cos 2I +\,\cos 2\mathit{\Omega}} + \right.\right.\\ \nonumber \\
\nonumber &\left.\left. +  2\,\ton{2 + 3\,e^2}\,\ton{-9 + 10\,\cos 2\mathit{\Omega}\,\sin^2 I}} -\right. \\ \nonumber \\
&\left. - 5\,e^2\,\ton{5 + 20\,\cos 2I + 7\,\cos 4I}\,\sin 2\omega\,\sin 2\mathit{\Omega}}, \\ \nonumber \\
\textsf{T}^{J_2\,\mathrm{dir}}_{yy} \nonumber &= -\rp{15}{512}\,{\mathcal{A}}_{J_2}\,\grf{e^2\,\qua{\ton{5 + 20\,\cos 2I +
         7\,\cos 4I}\,\cos 2\mathit{\Omega} - \right.\right. \\ \nonumber \\
\nonumber &\left.\left. -
      4\,\ton{5 + 7\,\cos 2I}\,\sin^2 I}\,\sin 2\omega +
   4\,\qua{e^2\,\ton{3\,\cos I + 5\,\cos 3I}\,\cos 2\omega +\right.\right. \\ \nonumber \\
&\left.\left. + 4\,\ton{2 + 3\,e^2}\,\cos I\,\sin^2 I}\,\sin 2\mathit{\Omega}}, \\ \nonumber \\
\textsf{T}^{J_2\,\mathrm{dir}}_{yz} \nonumber &= -\rp{15}{256}\,{\mathcal{A}}_{J_2}\,\qua{e^2\,\cos \mathit{\Omega}\,\ton{-22\,\sin 2I +
      7\,\sin 4I}\,\sin 2\omega +\right. \\ \nonumber \\
&\left. + 8\,e^2\,\cos 2\omega\,\ton{\sin I -\,\sin 3I}\,\sin \mathit{\Omega} +
   2\,\ton{2 + 3\,e^2}\,\ton{\sin I +
      5\,\sin 3I}\,\sin \mathit{\Omega}}, \\ \nonumber \\
\textsf{T}^{J_2\,\mathrm{dir}}_{zx} \nonumber \lb{zxdirJ2}&= -\rp{3}{256}\,{\mathcal{A}}_{J_2}\,\qua{-20\,e^2\,\cos 2\omega\,\cos \mathit{\Omega}\,\ton{\sin I - 7\,\sin 3I} - \right. \\ \nonumber \\
&\left. -14\,\ton{2 + 3\,e^2}\,\cos \mathit{\Omega}\,\ton{\sin I + 5\,\sin 3I} - 5\,e^2\,\ton{26\,\sin 2I + 7\,\sin 4I}\,\sin 2\omega\,\sin \mathit{\Omega}}, \\ \nonumber \\
\textsf{T}^{J_2\,\mathrm{dir}}_{zy} \nonumber &= -\rp{3}{256}\,{\mathcal{A}}_{J_2}\,\qua{5\,e^2\,\cos \mathit{\Omega}\,\ton{26\,\sin 2I +
      7\,\sin 4I}\,\sin 2\omega - \right. \\ \nonumber \\
&\left. - 20\,e^2\,\cos 2\omega\,\ton{\sin I - 7\,\sin 3I}\,\sin \mathit{\Omega} - 14\,\ton{2 + 3\,e^2}\,\ton{\sin I + 5\,\sin 3I}\,\sin \mathit{\Omega}}, \\ \nonumber \\
\textsf{T}^{J_2\,\mathrm{dir}}_{zz} \lb{zzdirJ2}&= -\rp{15}{64}\,{\mathcal{A}}_{J_2}\,e^2\,\ton{5 + 7\,\cos 2I}\,\sin^2 I\,\sin 2\omega.
\end{align}

It can be noted that \rfrs{xxdirJ2}{zzdirJ2} are independent of $f_0$, in agreement with Figure\,\ref{figura1}.
In the case of GP-B, \rfrs{zxdirJ2}{zzdirJ2} and \rfr{DEC} yield $\dot\delta = 5.1\,\mathrm{mas\,yr}^{-1}$; cfr. with  Figure\,\ref{figura1}.
For an exactly circular ($e=0$) and polar ($I=90^\circ$) orbit, by posing
\eqi
\mathcal{A}_{J_2}^{\ton{0}}\doteq \rp{\nk}{2}\,\ton{\rp{\mathcal{R}_\mathrm{s}}{a}}\,\ton{\rp{R}{a}}^2\,J_2,
\eqf
one has, from \rfrs{zxdirJ2}{zzdirJ2} and \rfr{DEC},
\eqi
\dert{\delta}{t} = -\rp{21}{16}\,\mathcal{A}_{J_2}^{\ton{0}}\,\cos\ton{\alpha-\mathit{\Omega}}.
\eqf
It agrees neither with Equation\,(53) of \citet{1970PhRvD...2.1428B}, which allows to obtain
\eqi
\dert{\delta}{t} = -\rp{9}{8}\,\mathcal{A}_{J_2}^{\ton{0}}\,\cos\ton{\alpha-\mathit{\Omega}},
\eqf
nor with $\ang{\Omega_G}$ of \citet[pag.\,153]{2003nlgd.conf..145A}, from which one gets
\eqi
\dert{\delta}{t} = -\rp{27}{16}\,\mathcal{A}_{J_2}^{\ton{0}}\,\cos\ton{\alpha-\mathit{\Omega}}.
\eqf
\subsection{The indirect effects}\lb{sub2.2}
This part treats what one may call the \virg{indirect}, or \virg{mixed}, effects arising from the precession of the orbit of the gyro caused by the oblateness of the primary. When applied to  \rfrs{TxxdS}{TzzdS}, they give rise to further components  of the gyro's spin rate of change of the order of $J_2\,c^{-2}$ which are to be added to the direct ones of Section\,\ref{sec3.2} in order to have the total expression of the 1pN spin rate due to $J_2$.
In turn, the calculation of the mixed effects can be split into two parts.

The first one, tagged in the following with the superscript ${\virg{\mathrm{mix\,I}}}$, consists of averaging  \rfrs{TxxdS}{TzzdS}, to be evaluated onto the unperturbed Keplerian ellipse, by means of \citep{1991ercm.book.....B,2014grav.book.....P}
\eqi
\widetilde{\dert{t}{f}} = \rp{r^4}{e\,\sqrt{\mu^3\,p}}\,\qua{-\cos f\,A_\mathrm{r} + \ton{1 + \rp{r}{p}}\,\sin f\,A_\mathrm{t}}\lb{dtdff}.
\eqf
It accounts for the instantaneous change of the line of the apsides; indeed, the orbital period $\Pb$ is just the time required by the test particle to return at the (moving) pericentre position along its path. In \rfr{dtdff},
\begin{align}
A_\mathrm{r} &=\bds A\bds\cdot\bds{\hat{r}}, \\ \nonumber \\
A_\mathrm{t} &= \bds A\bds\cdot\ton{\bds{\hat{h}}\bds\times\bds{\hat{r}}}
\end{align}
are the radial and transverse components, respectively, of the perturbing acceleration $\bds A$ inducing the slow variation of the otherwise fixed Keplerian ellipse. In the present case, it is
\eqi
{\bds A}_{J_2} = \rp{3\,\mu\,J_2\,R^2}{2\,r^4}\grf{ \qua{5\,\ton{\bds{\hat{k}}\bds\cdot\bds{\hat{r}}}^2 - 1}\bds{\hat{r}}   -2\,\ton{\bds{\hat{k}}\bds\cdot\bds{\hat{r}}}\,\bds{\hat{k}}  }.\lb{accel}
\eqf

The second part, labeled in the following with the superscript $\virg{\mathrm{mix\,II}}$, takes into account  the $J_2$-driven instantaneous changes experienced by the osculating Keplerian elements during an orbital revolution.
The mean variation of any of the spin components' rates $\mathrm{d}S^i/\mathrm{d}t,\,i=1,\,2,\,3$ over an orbital period occurring due to the aforementioned shifts can be worked out as
\eqi
\Delta \dot S^i = \rp{\nk}{2\pi}\,\sum^5_{j=1}\int_{f_0}^{f_0+2\pi}\grf{\derp{\ton{\mathrm{d}S^i/\mathrm{d}t}}{\phi_j}}_\mathrm{K}\Delta\phi_j\ton{f_0,\,f}\,\dert{t}{f}\,\mathrm{d}f,\,i=1,\,2,\,3\lb{grossa}
\eqf
where $f_0$ is the true anomaly at a referenced epoch $t_0$, and $\phi_1\doteq a,\,\phi_2\doteq e,\,\phi_3\doteq I,\,\phi_4\doteq\Omega,\phi_5\doteq\omega$. The instantaneous shifts of the Keplerian orbital elements
\eqi
\Delta\phi_j\ton{f_0,\,f} = \int_{f_0}^{f}\grf{\dert{\phi_j}{f^{'}}}_\mathrm{K}\,\mathrm{d}f^{'},\,j=1,\ldots 5,\lb{shi}
\eqf
are to be calculated in the usual perturbative way by integrating  the right-hand-sides of the corresponding Gauss equations \citep[e.g.][]{2011rcms.book.....K,2014grav.book.....P,SoffelHan19} from $f_0$ to a generic $f$. In the present case, the shifts of \rfr{shi} are due to the acceleration of \rfr{accel}.
The curly brackets $\grf{\ldots}_\mathrm{K}$ in \rfrs{grossa}{shi} denote that their content has to be evaluated onto the unperturbed Keplerian ellipse. In particular, $\mathrm{d}S^i/\mathrm{d}t,\,i=1,\,2,\,3$ are to be calculated by evaluating  \rfrs{TxxdS}{TzzdS} onto the Keplerian ellipse, while \rfr{deti} has to be used for the (Keplerian) expression of $\mathrm{d}t/\mathrm{d}f$ entering \rfr{grossa}.
%
%
%
%
\subsubsection{The impact of the motion of the  line of the apsides on the orbital period: the $\mathrm{I}$-type indirect effects}\lb{sec4.1}
Here, we display the analytical expressions of the average of \rfrs{TxxdS}{TzzdS} performed by means of \rfr{dtdff}. To avoid extremely cumbersome formulas, we show only those valid in an equatorial coordinate system.  They turn out to be
\begin{align}
\textsf{T}^{J_2\,\mathrm{mix\,I}}_{xx}\nonumber \lb{xxmix1} & =\,\rp{3}{1024}\,{\mathcal{A}}_{J_2}\,\ton{4\,e^2\,\sin^4 I\,\sin 4\omega +
    2\,\cos 2\mathit{\Omega}\,\grf{\qua{12\,\ton{7 + 2\,e^2}\,\cos 2I + \right.\right.\right. \\ \nonumber \\
\nonumber & \left.\left.\left. + \ton{11 + 6\,e^2}\,\ton{3 + \cos 4I}}\,\sin 2\omega +
 e^2\,\ton{3 + \cos 2I}\,\sin^2 I\,\sin 4\omega} +\right. \\ \nonumber \\
\nonumber & \left.+
    4\,\ton{8 + 3\,e^2}\,\ton{5\,\cos I + 3\,\cos 3 I}\,\cos 2\omega\,\sin 2\mathit{\Omega} +\right. \\ \nonumber \\
\nonumber & \left.+
    8\,\sin^2 I\,\grf{\qua{9 - 6\,e^2 + \ton{11 + 6\,e^2}\,\cos 2I}\,\sin 2\omega + \right.\right.\\ \nonumber \\
 &\left.\left. + \cos I\,\ton{20 + 7\,e^2 + e^2\,\cos 4\omega}\,\sin 2\mathit{\Omega}}},\,\\\,\nonumber\\
\textsf{T}^{J_2\,\mathrm{mix\,I}}_{xy}\nonumber & =-\rp{3}{512}\,{\mathcal{A}}_{J_2}\,\ton{12\,\cos 3 I\,\qua{18 +
\cos 2\omega\,\ton{1 - 2\,e^2 + 4\,\cos 2\mathit{\Omega}}} +\right. \\ \nonumber \\
\nonumber & \left. +
   4\,\cos I\,\grf{90 + 36\,e^2 +
\cos 2\omega\,\qua{-3 +
         6\,e^2 + \ton{20 + 3\,e^2 + 9\,e^2\,\cos 2I}\,\cos 2\mathit{\Omega}} + \right.\right. \\ \nonumber \\
\nonumber & \left.\left. + \qua{-54\,e^2 + \ton{20 + 7\,e^2 +
      e^2\,\cos 4\omega}\,\cos 2\mathit{\Omega}}\,\sin^2 I} - \right. \\ \nonumber \\
\nonumber & \left.- \grf{\qua{12\,\ton{7 + 2\,e^2}\,\cos 2I + \ton{11 + 6\,e^2}\,\ton{3 + \cos 4I}}\,\sin 2\omega +\right.\right. \\ \nonumber \\
& \left.\left. +
e^2\,\ton{3 + \cos 2I}\,\sin^2 I\,\sin 4\omega}\,\sin 2\mathit{\Omega}},\,\\\,\nonumber\\
\textsf{T}^{J_2\,\mathrm{mix\,I}}_{xz}\nonumber & = -\rp{3}{256}\,{\mathcal{A}}_{J_2}\,\ton{\cos\mathit{\Omega}\,\sin I\,\grf{92 + 25\,e^2 + \ton{4 + 30\,e^2}\,\cos 2\,\omega +\right.\right. \\ \nonumber \\
\nonumber & \left.\left. +
\cos 2I\,\qua{196 + 47\,e^2 - 6\,\ton{-10 + e^2}\,\cos 2\omega} +
      2\,e^2\,\cos 4\omega\,\sin^2 I} - \right. \\ \nonumber \\
& \left. - \grf{2\,\qua{5 + \ton{11 + 6\,e^2}\,\cos 2I}\,\sin  2I\,\sin 2\omega +
      2\,e^2\,\cos I\,\sin^3 I\,\sin 4\omega}\,\sin\,\mathit{\Omega}},\,\\\,\nonumber\\
\textsf{T}^{J_2\,\mathrm{mix\,I}}_{yx}\nonumber & = -\rp{3}{512}\,{\mathcal{A}}_{J_2}\,\ton{48\,\cos 3 I\,\cos 2\omega\,\cos 2\mathit{\Omega} +\right. \\ \nonumber \\
\nonumber &\left. +
   4\,\cos I\,\grf{-36\,\ton{4 +
   e^2} + \qua{54\,\ton{4 + e^2} + \ton{20 + 7\,e^2 +
      e^2\,\cos 4\omega}\,\cos 2\mathit{\Omega}}\,\sin^2 I +\right.\right. \\ \nonumber \\
\nonumber &\left.\left. +
\cos 2\omega\,\qua{\ton{20 + 3\,e^2 + 9\,e^2\,\cos 2I}\,\cos 2\mathit{\Omega} +
         12\,\ton{1 - 2\,e^2}\,\sin^2 I}} - \right. \\ \nonumber \\
\nonumber &\left. -\grf{\qua{12\,\ton{7 + 2\,e^2}\,\cos 2I + \ton{11 + 6\,e^2}\,\ton{3 + \cos 4I}}\,\sin 2\omega +\right.\right. \\ \nonumber \\
&\left.\left. +
e^2\,\ton{3 + \cos 2I}\,\sin^2 I\,\sin 4\omega}\,\sin 2\mathit{\Omega}},\,\\\,\nonumber\\
\textsf{T}^{J_2\,\mathrm{mix\,I}}_{yy}\nonumber & = -\rp{3}{512}\,{\mathcal{A}}_{J_2}\,\ton{-2\,e^2\,\sin^4 I\,\sin 4\omega +
\cos 2\mathit{\Omega}\,\grf{\qua{12\,\ton{7 + 2\,e^2}\,\cos 2I + \right.\right.\right. \\ \nonumber \\
\nonumber &\left.\left.\left. +\ton{11 + 6\,e^2}\,\ton{3 + \cos 4I}}\,\sin 2\omega +
e^2\,\ton{3 + \cos 2I}\,\sin^2 I\,\sin 4\omega} +\right. \\ \nonumber \\
\nonumber &\left. +
   2\,\ton{8 + 3\,e^2}\,\ton{5\,\cos I + 3\,\cos 3 I}\,\cos 2\omega\,\sin 2\mathit{\Omega} +
   4\,\sin^2 I\,\grf{-\qua{9 - 6\,e^2 + \right.\right.\right. \\ \nonumber \\
 &\left.\left.\left. + \ton{11 + 6\,e^2}\,\cos 2I}\,\sin 2\omega +
\cos I\,\ton{20 + 7\,e^2 + e^2\,\cos 4\omega}\,\sin 2\mathit{\Omega}}},\,\\\,\nonumber\\
\textsf{T}^{J_2\,\mathrm{mix\,I}}_{yz}\nonumber & = -\rp{3}{256}\,{\mathcal{A}}_{J_2}\,\ton{2\,\cos\mathit{\Omega}\,\grf{\qua{5 + \ton{11 +
            6\,e^2}\,\cos 2I}\,\sin  2I\,\sin 2\omega +\right.\right. \\ \nonumber \\
\nonumber &\left.\left. +
e^2\,\cos I\,\sin^3 I\,\sin 4\omega} + \sin I\,\grf{92 + 25\,e^2 + \ton{4 + 30\,e^2}\,\cos 2\omega +\right.\right. \\ \nonumber \\
&\left.\left. +
\cos 2I\,\qua{196 + 47\,e^2 - 6\,\ton{-10 + e^2}\,\cos 2\omega} +
      2\,e^2\,\cos 4\omega\,\sin^2 I}\,\sin\,\mathit{\Omega}},\,\\\,\nonumber\\
\textsf{T}^{J_2\,\mathrm{mix\,I}}_{zx}\nonumber \lb{zxmix1} &= -\rp{3}{256}\,{\mathcal{A}}_{J_2}\,\ton{\cos\mathit{\Omega}\,\sin I\,\grf{-52 - 11\,e^2 + 2\,\ton{14 - 9\,e^2}\,\cos 2\,\omega +\right.\right. \\ \nonumber \\
\nonumber &\left.\left. +
\cos 2I\,\qua{-236 - 61\,e^2 + 6\,\ton{6 + 7\,e^2}\,\cos 2\omega} +
      2\,e^2\,\cos 4\omega\,\sin^2 I} - \right. \\ \nonumber \\
&\left. - \grf{2\,\qua{5 + \ton{11 + 6\,e^2}\,\cos 2I}\,\sin  2I\,\sin 2\omega +
      2\,e^2\,\cos I\,\sin^3 I\,\sin 4\omega}\,\sin\,\mathit{\Omega}},\,\\\,\nonumber\\
\textsf{T}^{J_2\,\mathrm{mix\,I}}_{zy}\nonumber &= -\rp{3}{256}\,{\mathcal{A}}_{J_2}\,\sin I\,\ton{2\,\cos I\,\cos\mathit{\Omega}\,\grf{2\,\qua{5 + \ton{11 + 6\,e^2}
\cos 2I}\,\sin 2\omega + \right.\right. \\ \nonumber \\
\nonumber &\left.\left. + e^2\,\sin^2 I\,\sin 4\omega} + \grf{-52 -
      11\,e^2 + 2\,\ton{14 - 9\,e^2}\,\cos 2\omega +\right.\right. \\ \nonumber \\
 &\left.\left. +
\cos 2I\,\qua{-236 - 61\,e^2 + 6\,\ton{6 + 7\,e^2}\,\cos 2\omega} +
      2\,e^2\,\cos 4\omega\,\sin^2 I}\,\sin\,\mathit{\Omega}},\,\\\,\nonumber\\
\textsf{T}^{J_2\,\mathrm{mix\,I}}_{zz} \lb{zzmix1}&= -\rp{3}{128}\,{\mathcal{A}}_{J_2}\,\sin^2 I\,\grf{2\,\qua{1 + 6\,e^2 + \ton{11 + 6\,e^2}\,
\cos 2I}\,\sin 2\omega + e^2\,\sin^2 I\,\sin 4\omega}.
\end{align}
It can be noted that \rfrs{xxmix1}{zzmix1} are independent of $f_0$.
For an exactly circular and polar orbit, \rfrs{zxmix1}{zzmix1} and \rfr{DEC} yield
\eqi
\dert{\delta}{t} = \rp{3}{32}\,\mathcal{A}^{\ton{0}}_{J_2}\,\qua{(-23 + \cos 2\omega)\,\cos\ton{\alpha - \mathit{\Omega}} +
 5\,\sin 2\omega\,\tan\delta}.\lb{deldir}
\eqf
While, seemingly, \citet{2003nlgd.conf..145A} did not deal with the issue of the indirect effects at all, \citet{1970PhRvD...2.1428B} did partly so.
Their Equation\,(52) allows to infer
\eqi
\dert{\delta}{t} = \rp{3}{8}\,\mathcal{A}^{\ton{0}}_{J_2}\,\cos\ton{\alpha - \mathit{\Omega}},\lb{merda}
\eqf
which disagrees with \rfr{deldir}. However, since it is unclear how \citet{1970PhRvD...2.1428B} actually calculated their indirect precession, it is uncertain that \rfr{merda} can meaningfully be compared with \rfr{deldir}.
\subsubsection{The impact of the instantaneous shifts of the orbital elements during an orbital revolution: the $\mathrm{II}$-type indirect effects}\lb{sec4.2}
Here, we display the analytical expressions of the average of \rfrs{TxxdS}{TzzdS} calculated according to \rfrs{grossa}{shi}. Because of their exceptional cumbersomeness, we can only show  their limit for $e\rightarrow 0$ in an equatorial coordinate system.

One has
\begin{align}
\textsf{T}^{J_2\,\mathrm{mix\,II}}_{xx}\nonumber \lb{xxmix2} &= \rp{\mathcal{A}_{J_2}^{\ton{0}}}{512}\,\grf{
-48\,\cos^2 I\,\cos 2\mathit{\Omega}\,\sin I -
 3\,\qua{\ton{15 + 44\,\cos 2I + 5\,\cos 4I}\,\cos 2\mathit{\Omega} + \right.\right.\\ \nonumber \\
\nonumber &\left.\left. +  4\,\ton{7 + 5\,\cos 2I}\,\sin^2 I}\,\sin 2\omega +
 12\,\qua{\ton{7 + 20\,\cos 2I + 5\,\cos 4I}\,\cos 2\mathit{\Omega} + \right.\right.\\ \nonumber \\
\nonumber &\left.\left. +
 4\,\ton{3 + 5\,\cos 2I}\,\sin^2 I}\,\sin 2\ton{f_0+\omega} + \right.\\ \nonumber \\
\nonumber &\left. +  8\,\cos I\,\qua{3\,\sin 2\ton{I - \mathit{\Omega}} +\,\ton{-23 + 26\,\cos 2I -\right.\right.\right.\\ \nonumber \\
\nonumber &\left.\left.\left. -
 9\,\cos 2\ton{I -\omega} - 6\,\cos 2\omega +
 24\,\cos 2\ton{f_0 - I +\omega} - 9\,\cos 2\ton{I +\omega} +\right.\right.\right.\\ \nonumber \\
&\left.\left.\left. +
 24\,\cos 2\ton{f_0 + I +\omega}}\,\sin 2\mathit{\Omega}}
}, \\ \nonumber \\
\textsf{T}^{J_2\,\mathrm{mix\,II}}_{xy}\nonumber \lb{xymix2} &= \rp{\mathcal{A}_{J_2}^{\ton{0}}}{512}\,\qua{
4\,\ton{3\,\cos 3I\,\ton{-7 + 6\,\cos 2\omega}\,\cos 2\mathit{\Omega} +
\,\cos I\,\grf{6\,\cos\ton{2I - 2 \mathit{\Omega}} + \right.\right.\right.\\ \nonumber \\
\nonumber &\left.\left.\left. + \qua{5\,\ton{5 + 6\,\cos 2\omega} -
 2\,\cos 2I\,\ton{5 + 48\,\cos 2\ton{f_0+\omega}}}\,\cos 2\mathit{\Omega} -\right.\right.\right.\\ \nonumber \\
\nonumber &\left.\left.\left. -
 36\,\qua{3 +\,\ton{3\,\cos 2\omega + 8\,\cos 2\ton{f_0+\omega}}\,\sin^2 I}} -
 72\,\cos 2\ton{f_0+\omega}\,\sin I\,\sin 2I -\right.\right.\\ \nonumber \\
\nonumber &\left.\left. -
 81\,\csc I\,\sin 4I} -
 3\,\qua{16\,\cos^2 I\,\sin I +\,\ton{15 + 44\,\cos 2I + 5\,\cos 4I}\,\sin 2\omega -\right.\right.\\ \nonumber \\
&\left.\left. -
 4\,\ton{7 + 20\,\cos 2I + 5\,\cos 4I}\,\sin 2\ton{f_0+\omega}}\,\sin 2\mathit{\Omega}
},\\ \nonumber \\
\textsf{T}^{J_2\,\mathrm{mix\,II}}_{xz}\nonumber \lb{xzmix2} &= \rp{\mathcal{A}_{J_2}^{\ton{0}}}{256}\,\grf{
2\,\qua{-116 - 376\,\cos 2I + 45\,\cos 2\ton{I -\omega} - 42\,\cos 2\omega - 120\,\cos 2\ton{f_0+\omega} +\right.\right.\\ \nonumber \\
\nonumber &\left.\left. +
60\,\cos 2\ton{f_0 - I +\omega} + 45\,\cos 2\ton{I +\omega} +
60\,\cos 2\ton{f_0 + I +\omega}}\,\cos\mathit{\Omega}\,\sin I +\right.\\ \nonumber \\
\nonumber &\left. +
12\,\cos\ton{I - \mathit{\Omega}}\,\sin 2I -
3\,\qua{5\,\sin 4I\,\sin 2\omega +
6\,\sin 2I\,\ton{16\,\pi +\,\sin 2\omega} +\right.\right.\\ \nonumber \\
& \left.\left. +
8\,\cos I\,\sin^2 I\,\ton{1 +
20\,\sin I\,\sin 2\ton{f_0+\omega}}}\,\sin\mathit{\Omega}
},\\ \nonumber \\
\textsf{T}^{J_2\,\mathrm{mix\,II}}_{yx}\nonumber \lb{yxmix2} &= \rp{\mathcal{A}_{J_2}^{\ton{0}}}{512}\,\qua{
4\,\ton{3\,\cos 3I\,\ton{-7 + 6\,\cos 2\omega}\,\cos 2\mathit{\Omega} +
\,\cos I\,\grf{108 +
 6\,\cos\ton{2I - 2 \mathit{\Omega}} + \right.\right.\right.\\ \nonumber \\
\nonumber &\left.\left.\left. +  \qua{5\,\ton{5 + 6\,\cos 2\omega} -
 2\,\cos 2I\,\ton{5 + 48\,\cos 2\ton{f_0+\omega}}}\,\cos 2\mathit{\Omega} +\right.\right.\right.\\ \nonumber \\
\nonumber &\left.\left.\left. +
 36\,\ton{3\,\cos 2\omega + 8\,\cos 2\ton{f_0+\omega}}\,\sin^2 I} + \right.\right.\\ \nonumber \\
\nonumber &\left.\left. +  72\,\cos 2\ton{f_0+\omega}\,\sin I\,\sin 2I +
 81\,\csc I\,\sin 4I} -\right.\\ \nonumber \\
\nonumber &\left. -
 3\,\qua{16\,\cos^2 I\,\sin I +\,\ton{15 + 44\,\cos 2I + 5\,\cos 4I}\,\sin 2\omega -\right.\right.\\ \nonumber \\
&\left.\left. -
 4\,\ton{7 + 20\,\cos 2I + 5\,\cos 4I}\,\sin 2\ton{f_0+\omega}}\,\sin 2\mathit{\Omega}
}, \\ \nonumber \\
\textsf{T}^{J_2\,\mathrm{mix\,II}}_{yy}\nonumber \lb{yymix2} &= \rp{\mathcal{A}_{J_2}^{\ton{0}}}{512}\,\csc I\,\ton{
-\rp{3}{2}\,\qua{8\,\ton{7 + 5\,\cos 2I}\,\sin^3 I + \cos 2\mathit{\Omega}\,\ton{14\,\sin I - 39\,\sin 3I -\right.\right.\right.\\ \nonumber \\
\nonumber &\left.\left.\left. -
 5\,\sin 5I}}\,\sin 2\omega +
 6\,\grf{8\,\ton{3 + 5\,\cos 2I}\,\sin^3 I +
 \cos 2\mathit{\Omega}\,\qua{6\,\sin I -\right.\right.\right.\\ \nonumber \\
\nonumber &\left.\left.\left. -
 5\,\ton{3\,\sin 3I +\,\sin 5I}}}\,\sin 2\ton{f_0+\omega} +
 2\,\qua{2\,\ton{23 + 6\,\cos 2\omega}\,\sin 2I + \right.\right.\\ \nonumber \\
&\left.\left. + \ton{-23 + 18\,\cos 2\omega - 48\,\cos 2\ton{f_0+\omega}}\,\sin 4I}\,\sin 2\mathit{\Omega}
}, \\ \nonumber \\
\textsf{T}^{J_2\,\mathrm{mix\,II}}_{yz}\nonumber \lb{yzmix2} &= \rp{\mathcal{A}_{J_2}^{\ton{0}}}{128}\,\ton{
3\,\sin 2I\,\grf{\cos\mathit{\Omega}\,\qua{48\,\pi +
 2\,\sin I +\,\ton{3 + 5\,\cos 2I}\,\sin 2\omega + \right.\right.\right.\\ \nonumber \\
\nonumber &\left.\left.\left. +  40\,\sin^2 I\,\sin 2\ton{f_0+\omega}} -
 2\,\sin\ton{I - \mathit{\Omega}}} + \right.\\ \nonumber \\
\nonumber &\left. +  \ton{-116 - 376\,\cos 2I +
 45\,\cos 2\ton{I -\omega} -  \right.\right.\\ \nonumber \\
\nonumber &\left.\left. - 42\,\cos 2\omega -
 120\,\cos 2\ton{f_0+\omega} + 60\,\cos 2\ton{f_0 - I +\omega} +
 45\,\cos 2\ton{I +\omega} +  \right.\right.\\ \nonumber \\
&\left.\left. + 60\,\cos 2\ton{f_0 + I +\omega}}\,\sin I\,\sin\mathit{\Omega}
}, \\ \nonumber \\
\textsf{T}^{J_2\,\mathrm{mix\,II}}_{zx}\nonumber \lb{zxmix2} &= \rp{\mathcal{A}_{J_2}^{\ton{0}}}{128}\,\ton{\qua{100 + 272\,\cos 2I - 9\,\cos 2\ton{I -\omega} + 66\,\cos 2\omega + 24\,\cos 2\ton{f_0+\omega} - \right.\right.\\ \nonumber \\
\nonumber &\left.\left. -
 156\,\cos 2\ton{f_0 - I +\omega} - 9\,\cos 2\ton{I +\omega}  -  156\,\cos 2\ton{f_0 + I +\omega}}\,\cos\mathit{\Omega}\,\sin I + \right.\\ \nonumber \\
\nonumber &\left. +
 3\,\sin 2I\,\grf{2\,\cos\ton{I - \mathit{\Omega}} +\,\qua{48\,\pi -
 2\,\sin I -\,\ton{3 + 5\,\cos 2I}\,\sin 2\omega + \right.\right.\right.\\ \nonumber \\
&\left.\left.\left. +
 4\,\ton{7 + 5\,\cos 2I}\,\sin 2\ton{f_0+\omega}}\,\sin\mathit{\Omega}}
}, \\ \nonumber \\
\textsf{T}^{J_2\,\mathrm{mix\,II}}_{zy} \nonumber \lb{zymix2} &= -\rp{\mathcal{A}_{J_2}^{\ton{0}}}{128}\,\grf{
3\,\cos\mathit{\Omega}\,\qua{13\,\cos 3I\,\sin I\,\sin 2\omega + \sin 4I\,\ton{-9\,\sin 2\omega + 10\,\sin 2\ton{f_0+\omega}} + \right.\right. \\ \nonumber \\
\nonumber &\left.\left. +  \sin 2I\,\ton{48\,\pi - 2\,\sin I + 7\,\cos\omega\,\sin\omega +
 28\,\sin 2\ton{f_0+\omega}}} + 6\,\sin 2I\,\sin\ton{I - \mathit{\Omega}} + \right. \\ \nonumber \\
\nonumber & \left. + \qua{-100 - 272\,\cos 2I + 9\,\cos 2\ton{I -\omega} - 66\,\cos 2\omega - 24\,\cos 2\ton{f_0+\omega} + \right.\right. \\ \nonumber \\
&\left.\left. + 156\,\cos 2\ton{f_0 - I +\omega} + 9\,\cos 2\ton{I +\omega} + 156\,\cos 2\ton{f_0 + I +\omega}}\,\sin I\,\sin\mathit{\Omega}
}, \\ \nonumber \\
\textsf{T}^{J_2\,\mathrm{mix\,II}}_{zz}\nonumber \lb{zzmix2} &= -\rp{3\,\mathcal{A}_{J_2}^{\ton{0}}}{64}\,\sin^2 I\,\grf{
4\,\ton{3 + 5\,\cos 2I}\,\cos 2\omega\,\sin 2f_0 + \right.\\ \nonumber \\
&\left. + \qua{1 - 5\,\cos 2I + 4\,\cos 2f_0\,\ton{3 + 5\,\cos 2I}}\,\sin 2\omega
}.
\end{align}

The dependence of \rfrs{xxmix2}{zzmix2} on $f_0$ is apparent.
\subsection{The total (direct + mixed) spin precessions of the order of $\mathcal{O}\ton{J_2\,c^{-2}}$}\lb{totJ2}
The results of Sections \ref{sec3.2} to \ref{sec4.2} allow to obtain the total 1pN spin precession due to the oblateness of the primary. It is not possible to display them here in full due to their cumbersomeness.
As an example, for an exactly circular  and polar  orbit, we have
\begin{align}
\dert{\delta}{t} \nonumber \lb{megaDEC}&= \rp{\mathcal{A}^{\ton{0}}_{J_2}}{16}\,\qua{\ton{-77 + 12\,\cos 2\omega + 42\,\cos 2\ton{f_0+\omega}}\,\cos\ton{\alpha - \mathit{\Omega}} + \right.\\ \nonumber \\
&\left. +  3\,\ton{\sin 2\omega + 2\,\sin 2\ton{f_0+\omega}}\,\tan \delta}, \\ \nonumber \\
\ton{\dert{\alpha}{t}}^2 \nonumber \lb{megaRA}& =\rp{\ton{\mathcal{A}^{\ton{0}}_{J_2}}^2}{4096}\,\sec^2\delta\,\ton{
16\,\qua{\ton{-67 + 6\,\cos 2\omega + 30\,\cos 2\ton{f_0 + \omega}}\,\sin\delta + \right.\right.\\ \nonumber \\
\nonumber &\left.\left. + 3\,\cos\delta\,\cos\ton{\alpha-\mathit{\Omega}}\,\ton{\sin 2\omega + 2\,\sin 2\ton{f_0 + \omega}}}^2\,\sin^2\mathit{\Omega} +\right.\\ \nonumber \\
\nonumber &\left. + \grf{4\,\cos\mathit{\Omega}\,\qua{\ton{67 - 6\,\cos 2\omega - 30\,\cos 2\ton{f_0 + \omega}}\,\sin\delta - \right.\right.\right.\\ \nonumber \\
\nonumber &\left.\left.\left. - 3\,\cos\alpha\,\cos\delta\,\cos\mathit{\Omega}\,\ton{\sin 2\omega + 2\,\sin 2\ton{f_0 + \omega}}} - \right.\right.\\ \nonumber \\
\nonumber &\left.\left. - 6\,\cos\delta\,\sin\alpha\,\ton{\sin 2\omega + 2\,\sin 2\ton{f_0 + \omega}}\,\sin 2\mathit{\Omega}}^2 - \right.\\ \nonumber \\
\nonumber &\left. - 16\,\sin^2\delta\,\qua{\ton{-77 + 12\,\cos 2\omega + 42\,\cos 2\ton{f_0 + \omega}}\,\cos\ton{\alpha-\mathit{\Omega}} + \right.\right.\\ \nonumber \\
&\left.\left. + 3\,\ton{\sin 2\omega + 2\,\sin 2\ton{f_0 + \omega}}\,\tan\delta}^2
}.
\end{align}
If $\delta=0^\circ,\,\alpha=\mathit{\Omega}+180^\circ$, as for GP-B, \rfrs{megaDEC}{megaRA} reduce to.
\begin{align}
\dert\delta t \lb{minidelta}& = -\rp{\mathcal{A}^{\ton{0}}_{J_2}}{16}\,\qua{-77 + 12\,\cos 2\omega + 42\,\cos 2\ton{f_0+\omega}}, \\ \nonumber \\
\ton{\dert\alpha t}^2 \lb{minira}& = \rp{9}{256}\,\ton{\mathcal{A}^{\ton{0}}_{J_2}}^2\,\ton{\sin 2\omega + 2\,\sin 2\ton{f_0 + \omega}}^2.
\end{align}
From \rfrs{minidelta}{minira} it can be noted that, since the pericentre of a polar orbit, in general, does undergo a secular precession due to $J_2$ \citep{2005som..book.....C}, the shift of the spin's right ascension is, actually, a harmonic signal with half the period\footnote{For GP-B, it is $P_\omega= - 0.3\,\mathrm{yr}$.} $P_\omega$ of the pericentre, while the spin's declination experiences a genuine secular trend superimposed to a harmonic pattern with $P_\omega/2$.

\begin{figure}[htb!]
\begin{center}
\centerline{
\vbox{
\begin{tabular}{c}
\epsfysize= 8.0 cm\epsfbox{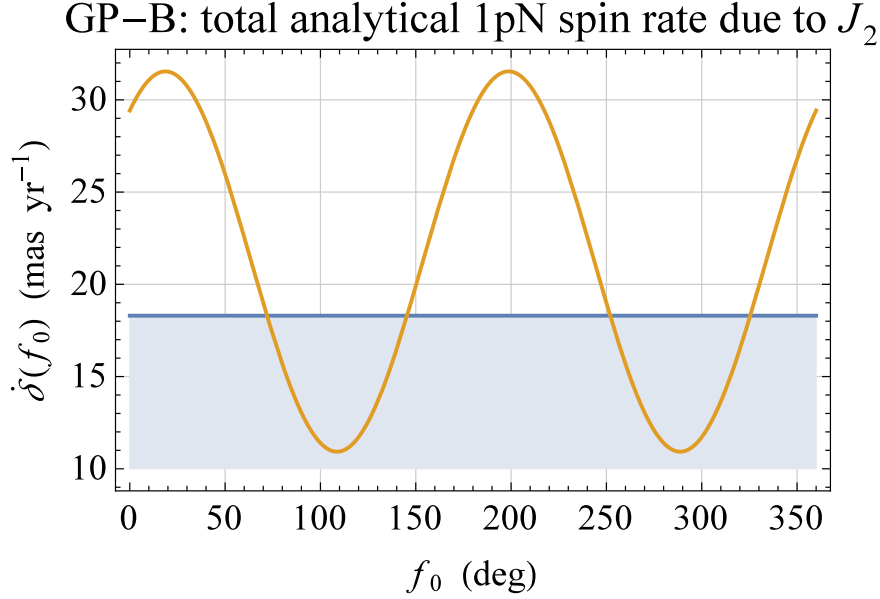}\\
\end{tabular}
}
}
\caption{
Total (direct + indirect) analytically computed 1pN $J_2$-induced rate of change $\dot\delta$ (yellow curve, in $\mathrm{mas\,yr}^{-1}$) of the declination $\delta$ of the spin axis of a gyroscope orbiting the oblate Earth as a function of the initial value $f_0$ of the true anomaly of the gyro's orbit.
We adopted the GP-B's orbital and spin configuration \citep{TabGPB} summarized in Table\,\ref{tavola1}, so that, essentially, the plot of \rfr{minidelta} is shown.  The shaded area, in light blue, is delimited by the GP-B's experimental mean uncertainty $\sigma^\mathrm{GP-B}_{\dot\delta}=18.3\,\mathrm{mas\,yr}^{-1}$ \citep{2011PhRvL.106v1101E,2015CQGra..32v4001E} in measuring the long-term rates of change of $\delta$. Cfr. with Figure\,\ref{figura2}.}\label{figura3}
\end{center}
\end{figure}

In the case of GP-B,  we plot its spin's declination precession as a function of $f_0$ in Figure\,\ref{figura3}.
It can be noted that the predicted rate is larger than $\upsigma_{\dot\delta}^\mathrm{GP-B}$ for $0^\circ\leq f_0\lesssim 70^\circ,\,150^\circ\lesssim f_0\lesssim 250^\circ,\,325^\circ\lesssim f_0\leq 360^\circ$, with peaks of more than $30\,\mathrm{mas\,yr}^{-1}$.
A comparison with Figure\,\ref{figura2} shows agreement between our analytical and numerical results up to a few $\mathrm{mas\,yr}^{-1}$.

We do not display the total GP-B's 1pN right ascension rate due to $J_2$ since  it turned out to be smaller than $2\,\mathrm{mas\,yr}^{-1}$, while the reported experimental accuracy in measuring $\dot\alpha$ is as large as $\upsigma^\mathrm{GP-B}_{\dot\alpha}=7.2\,\mathrm{mas\,yr}^{-1}$ \citep{2011PhRvL.106v1101E,2015CQGra..32v4001E}.
\section{The gravitomagnetic spin precession}\lb{gravimat}
The long-term gravitomagnetic spin precession induced by the proper angular momentum $\bds J$ of the primary can be analytically worked out by including \citep{2001rsgc.book.....R,2002NCimB.117..743R}
\eqi
g_{0i} = 2\rp{G\,\varepsilon_{ijk}\,J^j\,x^k}{c^3\,r^3},\,i=1,\,2,\,3,\lb{g0i}
\eqf
where
\eqi
\varepsilon_{ijk}  =\left\{
              \begin{array}{lll}
                +1 & \hbox{if $\ton{i,\,j,\,k}$ is $\ton{1,\,2,\,3}$,\,$\ton{2,\,3,\,1}$, or $\ton{3,\,1,\,2}$  } \\
                -1 & \hbox{if $\ton{i,\,j,\,k}$ is $\ton{3,\,2,\,1}$,\,$\ton{1,\,3,\,2}$, or $\ton{2,\,1,\,3}$ } \\
                0 & \hbox{if $i=j$, or $j=k$, or $k=i$}
              \end{array}
            \right.
\eqf
is the 3-dimensional Levi-Civita symbol \citep{levicivi},
in the spacetime metric tensor of \rfrs{g00}{gii}, and averaging  the resulting $J$-dependent part of the expansion of \rfr{spr} to the order of $\mathcal{O}\ton{c^{-2}}$ over a Keplerian ellipse. Smaller terms of the order of $\mathcal{O}\ton{J_2\,J\,c^{-2}}$, arising from using a $J_2$-driven precessing ellipse for the orbital average, will be neglected.

By defining the following dimensional amplitude having the dimension of reciprocal
time
\eqi
\mathcal{A}_\mathrm{gm}\doteq \rp{G\,J}{c^2\,a^3\,\ton{1-e^2}^{3/2}},
\eqf
one finally has
\begin{align}
\dert{S_x}{t} \lb{dsdxLT}  \nonumber & = -\rp{\mathcal{A}_\mathrm{gm}}{8}\,\qua{
-\ton{{\hat{J}}_y\, S_z + 2 S_y\, {\hat{J}}_z}\,\ton{1 + 3 \,\cos 2I} +\right.\\ \nonumber \\
\nonumber &+\left. 6\,{\hat{J}}_y\, S_z\,\cos 2\mathit{\Omega}\,\sin^2 I + 6\,\ton{S_y\, {\hat{J}}_y - S_z\, {\hat{J}}_z}\,\cos\mathit{\Omega}\,\sin 2I - 6\,{\hat{J}}_x\, S_y\,\sin 2I\,\sin\mathit{\Omega} -\right.\\ \nonumber \\
&-\left.6\,{\hat{J}}_x\, S_z\,\sin^2 I\,\sin 2\mathit{\Omega}}, \\ \nonumber \\
\dert{S_y}{t} \lb{dsdyLT} \nonumber & = -\rp{\mathcal{A}_\mathrm{gm}}{8}\,\qua{
\ton{{\hat{J}}_x\, S_z\, + 2 S_x\, {\hat{J}}_z}\,\ton{1 + 3 \,\cos 2I} + 6\,{\hat{J}}_x\, S_z\,\cos 2\mathit{\Omega}\,\sin^2 I -\right.\\ \nonumber \\
&-\left. 6\,S_x\, {\hat{J}}_y\,\cos\mathit{\Omega}\,\sin 2I +6\,\ton{S_x\, {\hat{J}}_x - S_z\, {\hat{J}}_z}\,\sin 2I\,\sin\mathit{\Omega} +
6\,{\hat{J}}_y\, S_z\,\sin^2 I\,\sin 2\mathit{\Omega}}, \\ \nonumber \\
\dert{S_z}{t} \lb{dsdzLT} \nonumber & = -\rp{\mathcal{A}_\mathrm{gm}}{8}\,\qua{
-{\hat{J}}_x\, S_y + S_x\, {\hat{J}}_y + \ton{-3 {\hat{J}}_x\, S_y + 3 S_x\, {\hat{J}}_y}\,\cos 2I -\right.\\ \nonumber\\
\nonumber &-\left. 6\,\ton{{\hat{J}}_x\, S_y\, + S_x\, {\hat{J}}_y}\,\cos 2\mathit{\Omega}\,\sin^2 I +
6\,{\hat{J}}_z\,\sin 2I \ton{S_x\,\cos\mathit{\Omega} + S_y\, \,\sin\mathit{\Omega}} +\right.\\ \nonumber \\
&+\left. 6\,\ton{S_x\, {\hat{J}}_x - S_y\, {\hat{J}}_y}\,\sin^2 I \,\sin 2\mathit{\Omega}},
\end{align}
where $\hat{J}_x,\,\hat{J}_y,\,\hat{J}_z$ are the components of the spin axis $\bds{\hat{J}}$ of the primary.
The gravitomagnetic averaged precessions  of \rfrs{dsdxLT}{dsdzLT} can be cast in the following  vectorial form
\eqi
\dert{\bds S}{t} = {\mathbf{\Omega}}_\mathrm{gm}\bds\times\bds S,\lb{megaLT}
\eqf
with
\eqi
{\mathbf{\Omega}}_\mathrm{gm} = \rp{\mathcal{A}_\mathrm{gm}}{2}\,\grf{
3\,\qua{\Jl\,\bds{\hat{l}} +\Jm\,\bds{\hat{m}}}-2\,\bds{\hat{J}}}.
\eqf
It can be noted that \rfr{megaLT} agrees with, e.g., Equation\,(10.146b) of \citet{2014grav.book.....P} for $\hat{J}_y=\hat{J}_y=0,\,\hat{J}_z=1$ and $e\rightarrow 0$. It is also in agreement with Equation\,(29) of \citet{1970PhRvD...2.1428B} for any orientation of $\bds{\hat{J}}$ and $e\neq 0$.

Let us adopt a coordinate system aligned with the primary's equatorial plane such that $\hat{J}_y=\hat{J}_y=0,\,\hat{J}_z=1$. According to \rfrs{DEC}{RA} and \rfrs{dsdxLT}{dsdzLT}, the gravitomagnetic spin precessions of $\delta,\,\alpha$ turn out to be
\begin{align}
\dert\delta t \lb{DECLT}& = -\rp{3}{4}\,\mathcal{A}_\mathrm{gm}\,\sin 2I\,\cos\ton{\alpha-\mathit{\Omega}}, \\ \nonumber \\
\ton{\dert\alpha t}^2 \lb{RALT}&= \rp{\mathcal{A}_\mathrm{gm}^2}{16}\,\ton{1 + 3\,\cos^2 I -3\,\sin^2 I + 3\,\sin 2I\,\tan\delta\,\sin\ton{\alpha-\mathit{\Omega}}}^2.
\end{align}
\section{Summary and conclusions}\lb{fine}
The quadrupole mass moment $J_2$ of a body affects, among other things, also the general relativistic precession of the spin of an orbiting gyroscope. We worked out it, to the 1pN level, both numerically and analytically by taking into account also the effect that the $J_2$-driven change of the gyro's orbit has on the the long-term spin rate itself. Indeed, limiting to averaging out the instantaneous $J_2$-dependent part of the spin precession onto a Keplerian orbit is not sufficient to correctly reproduce the total spin rate of change to the order of $\mathcal{O}\ton{J_2\,c^{-2}}$. Also the instantaneous Newtonian orbital shifts due to $J_2$  have to be taken into account when the average of the 1pN de Sitter-like instantaneous part of the spin precession is performed. The latter contribution introduces a dependence of the total averaged spin rate of the order of $\mathcal{O}\ton{J_2\,c^{-2}}$ on the initial orbital phase $f_0$. Such a feature was confirmed, among other things, also by the simultaneous numerical integrations of the equations for the parallel transport of the spin and of the geodesic equations of the gyro's motion that we performed by varying $f_0$.

We applied our results to the past GP-B mission in the field of Earth by finding a net precession of the declination of the spin axis which may be as large as $\simeq 30-40\,\mathrm{mas\,yr}^{-1}$. Since the reported error in measuring the GP-B's declination rate amounts to $18.3\,\mathrm{mas\,yr}^{-1}$, our result may prompt a reanalysis of the data in order to see if the effect we predicted could be detected.

For the sake of completeness, we analytically worked out, to the 1pN level, also the general expression of the gravitomagnetic spin precession induced by the proper angular momentum $\bds J$ of the central body.

Both our numerical and analytical methods hold for an arbitrary orientation of the body's symmetry axis and for a general orbital configuration of the gyro. As such, they can be extended also to other astronomical and astrophysical scenarios of interest like, e.g., other planets of our solar system, exoplanets close to their parent stars, stars orbiting galactic supermassive black holes, tight binaries hosting compact stellar corpses. It is hardly necessary to mention that, years ago, spacecraft-based missions were proposed to measure the angular momenta of Jupiter and the Sun by means of the gravitomagnetic Pugh-Schiff spin precessions.
\bibliography{J2c2spin}{}

\begin{thebibliography}{30}
\expandafter\ifx\csname natexlab\endcsname\relax\def\natexlab#1{#1}\fi

\bibitem[{{Adler} \& {Silbergleit}(2003)}]{2003nlgd.conf..145A}
{Adler} R.~J., {Silbergleit} A.~S., 2003, in Nonlinear Gravitodynamics. The
  Lense-Thirring Effect, {Ruffini} R.~J., {Sigismondi} C., eds., World
  Scientific, Singapore, pp. 145--154

\bibitem[{{Barker} \& {O'Connell}(1970)}]{1970PhRvD...2.1428B}
{Barker} B.~M., {O'Connell} R.~F., 1970, Phys. Rev. D, 2, 1428

\bibitem[{{Breakwell}(1988)}]{1988nznf.conf..685B}
{Breakwell} J.~V., 1988, in Near Zero: New Frontiers of Physics, {Fairbank}
  J.~D., {Deaver} Jr. B.~S., {Everitt} C.~W.~F., {Michelson} P.~F., eds., W.~H.
  Freeman, New York, pp. 685--690

\bibitem[{{Brumberg}(1991)}]{1991ercm.book.....B}
{Brumberg} V.~A., 1991, {Essential Relativistic Celestial Mechanics}. Adam
  Hilger, Bristol

\bibitem[{{Capderou}(2005)}]{2005som..book.....C}
{Capderou} M., 2005, {Satellites: Orbits and missions}. Springer, Berlin

\bibitem[{{de Sitter}(1916)}]{1916MNRAS..77..155D}
{de Sitter} W., 1916, MNRAS, 77, 155

\bibitem[{{Everitt}(1974)}]{Varenna74}
{Everitt} C.~W.~F., 1974, in Proceedings of the International School of Physics
  \virg{Enrico Fermi}. Course LVI. Experimental Gravitation, {Bertotti} B.,
  ed., Academic Press, New York and London, pp. 331--360

\bibitem[{{Everitt} {et~al}\mbox{.}(2001){Everitt}, {Buchman}, {Debra},
  {Keiser}, {Lockhart}, {Muhlfelder}, {Parkinson}, \&
  {Turneaure}}]{2001LNP...562...52E}
{Everitt} C.~W.~F., {Buchman} S., {Debra} D.~B., {Keiser} G.~M., {Lockhart}
  J.~M., {Muhlfelder} B., {Parkinson} B.~W., {Turneaure} J.~P., 2001, in
  Lecture Notes in Physics, Berlin Springer Verlag, Vol. 562, Gyros, Clocks,
  Interferometers ...: Testing Relativistic Gravity in Space, {L{\"a}mmerzahl}
  C., {Everitt} C.~W.~F., {Hehl} F.~W., eds., pp. 52--82

\bibitem[{{Everitt} {et~al}\mbox{.}(2011){Everitt}, {Debra}, {Parkinson},
  {Turneaure}, {Conklin}, {Heifetz}, {Keiser}, {Silbergleit}, {Holmes},
  {Kolodziejczak}, {Al-Meshari}, {Mester}, {Muhlfelder}, {Solomonik}, {Stahl},
  {Worden}, {Bencze}, {Buchman}, {Clarke}, {Al-Jadaan}, {Al-Jibreen}, {Li},
  {Lipa}, {Lockhart}, {Al-Suwaidan}, {Taber}, \& {Wang}}]{2011PhRvL.106v1101E}
{Everitt} C.~W.~F. {et~al.}, 2011, Phys. Rev. Lett., 106, 221101

\bibitem[{{Everitt} {et~al}\mbox{.}(2015){Everitt}, {Muhlfelder}, {DeBra},
  {Parkinson}, {Turneaure}, {Silbergleit}, {Acworth}, {Adams}, {Adler},
  {Bencze}, {Berberian}, {Bernier}, {Bower}, {Brumley}, {Buchman}, {Burns},
  {Clarke}, {Conklin}, {Eglington}, {Green}, {Gutt}, {Gwo}, {Hanuschak}, {He},
  {Heifetz}, {Hipkins}, {Holmes}, {Kahn}, {Keiser}, {Kozaczuk}, {Langenstein},
  {Li}, {Lipa}, {Lockhart}, {Luo}, {Mandel}, {Marcelja}, {Mester}, {Ndili},
  {Ohshima}, {Overduin}, {Salomon}, {Santiago}, {Shestople}, {Solomonik},
  {Stahl}, {Taber}, {Van Patten}, {Wang}, {Wade}, {Worden}, {Bartel}, {Herman},
  {Lebach}, {Ratner}, {Ransom}, {Shapiro}, {Small}, {Stroozas}, {Geveden},
  {Goebel}, {Horack}, {Kolodziejczak}, {Lyons}, {Olivier}, {Peters}, {Smith},
  {Till}, {Wooten}, {Reeve}, {Anderson}, {Bennett}, {Burns}, {Dougherty},
  {Dulgov}, {Frank}, {Huff}, {Katz}, {Kirschenbaum}, {Mason}, {Murray},
  {Parmley}, {Ratner}, {Reynolds}, {Rittmuller}, {Schweiger}, {Shehata},
  {Triebes}, {VandenBeukel}, {Vassar}, {Al-Saud}, {Al-Jadaan}, {Al-Jibreen},
  {Al-Meshari}, \& {Al-Suwaidan}}]{2015CQGra..32v4001E}
{Everitt} C.~W.~F. {et~al.}, 2015, Classical Quant. Grav., 32, 224001

\bibitem[{{Fokker}(1920)}]{1921KNAB...23..729F}
{Fokker} A.~D., 1920, Versl. Kon. Ak. Wet., 29, 611

\bibitem[{{Haas} \& {Ross}(1975)}]{1975Ap&SS..32....3H}
{Haas} M.~R., {Ross} D.~K., 1975, Astrophys. Space Sci., 32, 3

\bibitem[{{Iorio}(2015)}]{2015IJMPD..2450067I}
{Iorio} L., 2015, Int. J. Mod. Phys. D, 24, 1550067

\bibitem[{{Iorio}(2019)}]{2019Univ....5..165I}
{Iorio} L., 2019, Universe, 5, 165

\bibitem[{{Kahn}(2007)}]{TabGPB}
{Kahn} R., 2007, in Gravity Probe B-Post Flight Analysis$\cdot$Final Report,
  Stanford University, Stanford, pp. 461--464

\bibitem[{{Kopeikin}, {Efroimsky} \& {Kaplan}(2011){Kopeikin}, {Efroimsky}, \&
  {Kaplan}}]{2011rcms.book.....K}
{Kopeikin} S., {Efroimsky} M., {Kaplan} G., 2011, {Relativistic Celestial
  Mechanics of the Solar System}. Weinheim: Wiley-VCH

\bibitem[{{Misner}, {Thorne} \& {Wheeler}(2017){Misner}, {Thorne}, \&
  {Wheeler}}]{2017grav.book.....M}
{Misner} C.~W., {Thorne} K.~S., {Wheeler} J.~A., 2017, {Gravitation}. Princeton
  University Press, Princeton

\bibitem[{{O'Connell}(1969)}]{1969Ap&SS...4..119O}
{O'Connell} R.~F., 1969, Astrophys. Space Sci., 4, 119

\bibitem[{{Ohanian} \& {Ruffini}(2013)}]{2013grsp.book.....O}
{Ohanian} H., {Ruffini} R., 2013, {Gravitation and Spacetime. Third Edition}.
  Cambridge University Press, Cambridge

\bibitem[{{Poisson} \& {Will}(2014)}]{2014grav.book.....P}
{Poisson} E., {Will} C.~M., 2014, {Gravity}. Cambridge: Cambridge Univ. Press

\bibitem[{{Pugh}(1959)}]{Pugh59}
{Pugh} G., 1959, {Proposal for a Satellite Test of the Coriolis Prediction of
  General Relativity}. Research Memorandum~11, Weapons Systems Evaluation
  Group, The Pentagon, Washington D.C.

\bibitem[{{Rindler}(2001)}]{2001rsgc.book.....R}
{Rindler} W., 2001, {Relativity: special, general, and cosmological}. Oxford
  University Press, Oxford, UK

\bibitem[{{Ruggiero} \& {Tartaglia}(2002)}]{2002NCimB.117..743R}
{Ruggiero} M.~L., {Tartaglia} A., 2002, Nuovo Cimento B, 117, 743

\bibitem[{{Schiff}(1960)}]{Schiff60}
{Schiff} L., 1960, Phys. Rev. Lett., 4, 215

\bibitem[{{Soffel} {et~al}\mbox{.}(1987){Soffel}, {Wirrer}, {Schastok},
  {Ruder}, \& {Schneider}}]{1988CeMec..42...81S}
{Soffel} M., {Wirrer} R., {Schastok} J., {Ruder} H., {Schneider} M., 1987,
  Celest. Mech. Dyn. Astr., 42, 81

\bibitem[{{Soffel}(1989)}]{Sof89}
{Soffel} M.~H., 1989, Relativity in Astrometry, Celestial Mechanics and
  Geodesy. Springer, Heidelberg

\bibitem[{{Soffel} \& {Han}(2019)}]{SoffelHan19}
{Soffel} M.~H., {Han} W.-B., 2019, {Applied General Relativity}, {Astronomy and
  Astrophysics Library}. Springer Nature Switzerland, Cham

\bibitem[{{Tyldesley}(1975)}]{levicivi}
{Tyldesley} J.~R., 1975, {An introduction to tensor analysis for engineers and
  applied scientists}. Longman, London

\bibitem[{{Will}(2018)}]{2018tegp.book.....W}
{Will} C.~M., 2018, {Theory and Experiment in Gravitational Physics. Second
  Edition}. Cambridge University Press, Cambridge

\bibitem[{{Zee}(2013)}]{2013Zee}
{Zee} A., 2013, {Einstein Gravity in a Nutshell}. Princeton University Press,
  Princeton

\end{thebibliography}

\end{document}